\documentclass[12pt]{article}
\usepackage[pdftex]{graphicx}        
\usepackage{enumerate}
\usepackage{natbib}
\usepackage{url} % not crucial - just used below for the URL 
\usepackage{amsmath,amsfonts,enumerate,xcolor,verbatim,booktabs,amssymb}
\usepackage{algorithm}
\usepackage{microtype}
\usepackage{xspace}
\usepackage{mathtools}
\usepackage[normalem]{ulem} 
\usepackage{times}
\usepackage{xr}
\usepackage[english]{babel}
\usepackage[figuresright]{rotating}

\newcommand{\blind}{1}

\renewcommand{\vec}[1]{\boldsymbol{#1}}
\newcommand{\lambdavec}{\vec\lambda}

\newcommand{\gammavec}{\vec\gamma}
\newcommand{\Wvec}{\vec{W}}
\newcommand{\cvec}{\vec{c}}

\newcommand{\xvec}{\vec{x}}

\newcommand{\indep}{\overset{indep}{\sim }}

\newcommand{\U}{\mathcal{P}}
\newcommand{\Ga}{\text{Ga}}
\newcommand{\Exp}{\text{Exp}}
\newcommand{\DP}{\text{DP}}
\newcommand{\D}{\mathcal{D}}

\begin{document}

\def\spacingset#1{\renewcommand{\baselinestretch}%
{#1}\small\normalsize} \spacingset{1}

%%%%%%%%%%%%%%%%%%%%%%%%%%%%%%%%%%%%%%%%%%%%%%%%%%%%%%%%%%%%%%%%%%%%%%%%%%%%%%

\if1\blind
{
  \title{\bf Revealing subgroup structure in ranked data using a Bayesian WAND}
%  \title{\bf Clustering ranked data using a Bayesian WAND}
  \author{S. R. Johnson%\thanks{The authors gratefully acknowledge
%\textit{please remember to list all relevant funding sources in the
%unblinded version}}
\hspace{.2cm}\\
    School of Mathematics, Statistics \& Physics, Newcastle University,\\ Newcastle upon Tyne, NE1 7RU, UK\\
    D. A. Henderson\\
     School of Mathematics, Statistics \& Physics, Newcastle University,\\ Newcastle upon Tyne, NE1 7RU, UK
    and \\
    R. J. Boys\\
     School of Mathematics, Statistics \& Physics, Newcastle University,\\ Newcastle upon Tyne, NE1 7RU, UK}
  \maketitle
} \fi

\if0\blind
{
  \bigskip
  \bigskip
  \bigskip
  \begin{center}
    {\LARGE\bf Revealing subgroup structure in ranked data using a Bayesian WAND}
\end{center}
  \medskip
} \fi

\bigskip
\begin{abstract}
  Ranked data arise in many areas of application ranging from the
  ranking of up-regulated genes for cancer to the ranking of academic
  statistics journals. Complications can arise when rankers do not
  report a full ranking of all entities; for example, they might only
  report their top--$M$ ranked entities after seeing some or all
  entities. It can also be useful to know whether rankers are equally
  informative, and whether some entities are effectively judged to be
  exchangeable. When there is important subgroup structure in the
  data, summaries such as aggregate (overall) rankings can be
  misleading. In this paper we propose a flexible Bayesian
  nonparametric model for identifying heterogeneous structure and
  ranker reliability in ranked data.  The model is a Weighted Adapted
  Nested Dirichlet (WAND) process mixture of Plackett-Luce models and
  inference proceeds through a simple and efficient Gibbs sampling
  scheme for posterior sampling.  The richness of information in the
  posterior distribution allows us to infer many details of the
  structure both between ranker groups and between entity groups
  (within ranker groups), in contrast to many other (Bayesian)
  analyses. We also examine how posterior predictive checks can be
  used to identify lack of model fit.  The methodology is illustrated
  using several simulation studies and real data examples.
\end{abstract}

\noindent%
\vfill

\newpage
\spacingset{1.45} % DON'T change the spacing!
\section{Introduction}
\label{sec:introduction}

Ranked data are central to many applications in science, social
  science, medicine and engineering and arise when $n$
\textit{rankers} (individuals) provide a ranking or ordering for a set
of $K$ \textit{entities} subject to some criterion.  A ranking or
ordering is therefore equivalent to a specific permutation of a set
and this perspective has proved useful for the development of models
for ranked data. There are many types of models for ranked data
  including parametric, stagewise and distance-based models; see
  overviews by \cite{Marden95} and \cite{AlvoY14}.  
\begin{comment}
The majority of
these models rely on strong assumptions about the homogeneity of
ranked data with the intention of determining an overall
  consensus ranking.  However, as we shall see, ranked data is often
  not homogeneous.  Distinguishing between homogeneous and
  heterogeneous data is important as it can be misleading to summarise
  a heterogeneous population as if its members held (roughly) the same
  consensus view. Our aim in this paper is to explore any subgroup
  structure within a set of ranked data and provide separate
  consensus rankings for each subgroup.
\end{comment}
Our work builds on the popular Plackett-Luce model
  \citep{luce1959indicidual,plackett1975analysis}. This model is an
extension to multiple comparison (ranked) data of the model for paired
comparisons proposed by \cite{bradley1952rank}. It caters for
commonplace complete and partial rankings and an extension also allows
it to deal with another increasingly popular type of ranking, namely
top--$M$ rankings; definitions of these various types of ranking are
given in Section~\ref{sec:model_rank_process}.

A standard assumption in the majority of the literature in this area
is that the data come from a homogeneous group of rankers, in which
each ranker only has fairly minor differences from an overall
consensus view. In general this assumption is not plausible and so the
main focus of this paper is to uncover any latent group structures
that exist within ranked data.  We aim to identify homogeneous groups
of individuals who share similar beliefs along with discovering how
some, or indeed all, of these groups may struggle to distinguish
between certain entities.  We also extend the model further to allow
for individuals who may be significantly more informed about the
entities they are ranking, the idea being that the opinions of such
individuals should carry more weight. Ranker reliability is introduced
into the model via ranker-specific latent binary indicators within the
Plackett-Luce likelihood; we refer to the resulting model as the
weighted Plackett-Luce model. In related work, \cite{deng2014rankagg}
used ranker reliability as part of their BARD (Bayesian Aggregation of
Ranked Data) solution to the rank aggregation problem and presented an
example involving the ranking of NBA teams which we revisit in
Section~\ref{sec:NBA_study}.

Appealing to mixtures to model potentially heterogeneous populations
is commonplace.  \cite{GormleyM08jasa,GormleyM08aas,gormley2009grade}
and \cite{MollicaT14,mollica2015bayesian} propose finite mixtures of
Plackett-Luce and related models to allow for different preferences
between rankers.  This approach was also taken by
\cite{vitelli2015prob}, however they adopted a distance-based model by
\cite{mallows1957non} rather than the Plackett-Luce model. Here we aim
to increase modelling flexibility by appealing to Bayesian
nonparametrics~\citep{HjortHMW10}. Specifically, we adopt an infinite
mixture of weighted Plackett-Luce models by means of a Dirichlet
process prior, thus allowing for full generality and allowing the
number of mixture components to be inferred from the data. Furthermore
we aim to investigate the latent grouping of entities.  We propose
that, within each group of rankers, there may be some underlying
belief that some of the entities are homogeneous.  That is to say, a
certain group of rankers may not be able to distinguish between all of
the entities and therefore we should allow for some to be judged as
equivalent.  This aspect is somewhat less explored within the
literature and approaches that allow for this have been limited to
distance based models \citep{marden1992use} and to the best of our
knowledge no such method exists for parametric models.  To allow for
the entity clustering within ranker groups we introduce a further set
of Dirichlet processes; we refer to these as the ``low level''
Dirichlet processes and there shall be a unique one for each of the
ranker clusters.  This results in an infinite set of ``skill
parameters'' for each ranker group from which $K$ are sampled.  The
construction is similar to that of the Nested Dirichlet Process (NDP)
\citep{rodriguez2008nested}, however adaptations to the NDP are
required in order to handle ranked data, and these are described in
Section~\ref{sec:andp}. Our Adapted Nested Dirichlet process (ANDP)
construction differs from the nonparametric Bayesian model proposed by
\cite{caron2014bayesiancluster} in that they focus solely on top--$M$
rankings and allow for a potentially infinite number of ``unobserved
entities'' (which have not yet appeared within any top--$M$ ranking)
to appear within a future ranking.  Our Adapted Nested Dirichlet
process prior, on the other hand, considers the situation where the
number of entities from which a ranking is produced is finite and
known, but utilises the nonparametric nature of the prior to
facilitate the entity grouping structure.

Our full model specification, which we call WAND (for Weighted Adapted
Nested Dirichlet process), provides a highly flexible framework for
analysing heterogeneous ranked data: the ANDP prior allows us to
cluster both rankers and entities and the weighted Plackett-Luce
likelihood allows for variable ranker reliability. Furthermore we
adopt a Bayesian approach to inference and propose an efficient Markov
chain Monte Carlo (MCMC) scheme for sampling from the posterior
distribution of the model parameters. The first step in the
construction of the computational scheme is to introduce an
appropriate version of the latent variables proposed by
\cite{caron2012efficient} for inference in the Plackett-Luce model.
For their suggested choice of prior distribution these latent variables
enable a semi-conjugate update for our skill parameters as part of a
Gibbs sampling algorithm.  To further increase computational
efficiency we appeal to marginal sampling methods for infinite
mixtures; namely Algorithm~8 of \cite{neal2000markov}.  We require a
nested version of this algorithm; the top level to sample the ranker
cluster allocations followed by the lower level which is used to
sample entity cluster allocations within each ranker cluster.
Marginal methods also have the additional appeal of integrating out
the infinite-dimensional aspect of the distribution and thus avoid any
approximations (and problems with label switching;
\cite{papa2008retrospective,hastie2015}) which would be obtained if we
implemented an algorithm based on the stick-breaking
construction~\citep{ishwaran2001gibbs}.

The remainder of the paper is structured as follows.  In
Section~\ref{sec:model_rank_process} we outline the Plackett-Luce
model and describe modifications for various types of rankings
(complete, partial, top--$M$). We also describe the binary latent
variable structure for modelling ranker reliability.
Sections~\ref{sec:clustering} and~\ref{sec:andp} describe the two-way
nested cluster structure and our proposed ANDP prior for dealing with
this structure. The weighted Plackett-Luce likelihood and the ANDP
prior are combined to give the WAND (Weighted Adapted Nested Dirichlet
process) and details of the prior specification, latent variables and
the Gibbs sampling algorithm for posterior sampling are given in
Section~\ref{sec:wand}. Two simulation studies illustrating the use of
the WAND model in various scenarios are given in
Section~\ref{sec:sim_studies} of the Supplementary materials and are
summaried briefly in Section~\ref{sec:wand} of this paper. This
section also describes a way to diagnose lack of model fit by using
posterior predictive checks.  Section~\ref{sec:data} in the paper
illustrates the use of the WAND model on two real datasets and
compares its results to those using other methods.
Section~\ref{sec:conc} offers some conclusions.

\section{Model for the ranking process}
\label{sec:model_rank_process}

We assume our data (rankings) are observations from the Plackett-Luce
model \citep{luce1959indicidual,plackett1975analysis}. We define the
set $\mathcal{K} = \{1,\dots,K\}$ which contains $K = |\mathcal{K}|$
entities to be ranked. All entities have a ``skill rating'' $\lambda_k
> 0$ for $k = 1,\dots,K$; let
$\vec{\lambda}=(\lambda_1,\ldots,\lambda_K)$.  Each ranking need not
contain every entity; let $n_i \leq K$ be the number of entities
contained within ranking~$i$. Thus, a typical observation from this
model is $\vec{x}_i = \{x_{i1},\dots,x_{in_i}\}$, where $x_{ij}$
denotes the entity ranked in position $j$ in the $i$th ranking. The
probability of such an observation under the Plackett-Luce model is
\begin{align}
\label{eqn:pl_model}
\Pr(\vec{X_i} = \vec{x_i} | \vec{\lambda}) &= \prod_{j=1}^{n_i}
\frac{\lambda_{x_{ij}}}{\sum_{m=j}^{n_i} \lambda_{x_{im}}}.
\end{align}
The Plackett-Luce probability can be understood through ideas of
  sampling without replacement, with the probability of choosing an
  entity being proportional to its $\lambda$-value.  Consider the
  ranking $\vec{x}=(4,3,2,1)$. The first term in
  $\eqref{eqn:pl_model}$ is the probability of choosing entity~4 from
  all four entities; the second term is the probability of choosing
  entity~3 from the remaining entities, and so on. Thus the
  $\lambda$-parameters influence the likely position of entities within
  rankings, and entities with larger values are more likely to be given
  a position higher up the ranking.

Suppose we have $n$ rankings. Let the data $\vec{x}_i$ for
$i=1,\dots,n$ be collectively denoted $\D$. The likelihood is
\[
\pi(\D|\vec{\lambda}) = \prod_{i=1}^n \prod_{j=1}^{n_i} \frac{\lambda_{x_{ij}}}{\sum_{m=j}^{n_i} \lambda_{x_{im}}},
\]
which is invariant to scalar multiplication of the parameters
$\lambdavec$.  This leads to a parameter identifiability issue and
potential mixing problems for MCMC algorithms.  The identifiability
issue is not of great concern as the parameters can be normalised as
required.  It is, however, desirable to resolve the mixing issue and
this can be achieved through suitable rescaling
\citep{caron2012efficient}; details of which are provided within
Section~\ref{sec:rescaling}.

A limitation of model~\eqref{eqn:pl_model} is that the probability is
only suitably defined for certain types of ranking.  The model
requires each ranker to report a position for each of the entities
they consider.  This allows for two types of ranking: (a)
\textit{complete rankings}, in which a ranker considers and ranks all
possible entities and (b) \textit{partial rankings}, in which a ranker
considers a subset of all the entities and reports back a position for
each, and so $n_i < K$ in this scenario. Note that this model can also
deal with paired comparisons as these are \textit{partial} rankings
with $n_i=2$. Two further popular types of ranking are
\textit{top--$M$ complete} and \textit{top--$M$ partial rankings}. A
top--$M$ complete ranking is obtained when an individual considers all
$K$~entities but only reports a ranking for those entities they
consider to be positioned 1 to $M$. In contrast, a top--$M$ partial
ranking is one obtained from an individual who only considers $K_i <
K$ entities and reports which of these entities they judge to be in
positions 1 to $M$.  The current model does not fully capture the
information in data of this type.  If we naively chose to include
top--$M$ rankings with $n_i = M$ in (\ref{eqn:pl_model}) then the
entity in position $M$ is treated as if it was ranked last.  The
additional information provided by the ranker that the entities not
given a position are ranked somewhere in positions $M+1$, $M+2$, and
so on, is therefore neglected. We can allow for these additional
ranking types by modifying the Plackett-Luce model. Recall that the
set of all entities is denoted by $\mathcal{K} = \{1,\dots,K\}$.  Now
suppose ranker $i$ considers $K_i \leq K$ entities and that the set of
these entities is denoted by $\mathcal{K}_i \subseteq \mathcal{K}$.
Further we introduce the set $\mathcal{U}_i = \mathcal{K}_i \setminus
\{ \vec{x}_i \}$ as the collection of entities considered by
ranker~$i$ that did not feature in their ranking, that is,
$\mathcal{U}_i$ contains entities that are somewhere in positions
$n_i+1$, $n_i+2$ and so on.  The Plackett-Luce model can deal with
these additional ranking types if it is modified to
\begin{equation}
\label{eqn:pl_modified_rankings} 
\Pr(\vec{X_i} = \vec{x_i} | \vec{\lambda}) 
= \prod_{j=1}^{n_i} \frac{\lambda_{x_{ij}}}
{\sum_{m=j}^{n_i} \lambda_{x_{im}} + \sum_{m \in \mathcal{U}_i} \lambda_m}.
\end{equation}
Note that if we have only \textit{complete} or \textit{partial}
rankings then the $\mathcal{U}_i = \varnothing$ and we recover the
standard Plackett-Luce model~\eqref{eqn:pl_model}.

Model (\ref{eqn:pl_modified_rankings}) above still relies on the
implicit assumption that each ranking is equally informative, that is,
each ranking is equally representative of the views taken by the
population. This rather strong assumption can be false such as when
some rankers are (substantially) more informed about the entities than
others. We choose to model such (potential) heterogeneity between
rankings via a mixture model with two components: one for the
informative rankings and the other for uninformative rankings. This is
similar in spirit to the mixture model formulation of \cite{MurphyM03}
in which a specific mixture component is used to model ``noise''
rankings. However, we implement our mixture model in a different way.
We introduce a latent binary indicator variable~$W_i$ for each ranking
($i=1,\dots,n$) with $W_i=1$ if ranking $i$ is informative and $W_i=0$
if it is not.  We incorporate informativeness by modifying the
Plackett-Luce model (\ref{eqn:pl_modified_rankings}) so that the
probability of the $i$th observed ranking is
\begin{equation}
\label{eqn:plprob_weights}
\Pr(\vec{X}_i=\vec{x}_i|\lambdavec,W_i=w_i)  
= \prod_{j=1}^{n_i}\frac{\lambda_{x_{ij}}^{w_i}}{\sum_{m=j}^{n_i}\lambda_{x_{im}}^{w_i} + \sum_{m \in \mathcal{U}_i} \lambda_m ^{w_i}}. \\
\end{equation}
Note that $\Pr(\vec{X}_i=\vec{x}_i|\lambdavec,W_i=0) =
(K_i-n_i)!/K_i!$, that is, the reciprocal of the number of ordered
permutations of $n_i$ entities from a set of $K_i$. Therefore $W_i=0$
corresponds to there being no information in ranking $i$; essentially
ranker $i$ has picked $\vec{x}_i$ uniformly at random from all
possible rankings of $n_i$ out of $K_i$ entities.  Otherwise
ranker~$i$ is deemed as informative ($W_i=1$) and the probability of
the observed ranking is as in \eqref{eqn:pl_modified_rankings}.  We
use the notation $\vec{X_i}|\lambdavec, w_i \sim
\text{PL}_\text{W}(\lambdavec, w_i)$ to denote that the probability of
observation/ranking $i$ is given by (\ref{eqn:plprob_weights}) and
call this model the weighted Plackett-Luce model. We note in passing
that, although allowing the weights~$w_i$ to be continuous, say within
$(0,1)$, is appealing, it does lead to identifiability problems for
$(\lambdavec,\Wvec)$ as multiple values give the same weighted
Plackett-Luce likelihood.  Also the choice of binary $W_i$ has the
benefit that $W_i=1$ recovers the standard Plackett-Luce distribution
and $W_i=0$ is meaningful in that it represents a uniform ranking
distribution.

\section{Clustering}
\label{sec:clustering}
Until now we have assumed that the dataset can be adequately
summarised by using a unique parameter vector $\lambdavec$.
Furthermore we have also assumed that each value $\lambda_j$
($j=1,\dots,K$) is unique.  We now suppose that there may be groups of
rankers, each of which has their own beliefs about the true ranking.
We do this by associating each ranker with a particular parameter
vector, say one of $\lambdavec_1,\ldots,\lambdavec_{N^r}$, and allow
rankers to share the same beliefs about entities by having the same
parameter vector. We will also allow for the possibility that ranker
groups are not able to distinguish between groups of entities by
allowing entities to be tied in strength ($\lambda$-value) so that
each parameter vector~$\lambdavec_i$ can have multiple entries of the
same value. We shall appeal to Bayesian nonparametric clustering
methods in order to implement this structure; namely Dirichlet
processes.

We use the notation $G \sim \DP(\alpha, G_0)$ to denote that $G$ is a
Dirichlet process, where $\alpha$ and $G_0$ are the concentration
parameter and base distribution respectively.  Each Dirichlet process
has a corresponding stick-breaking representation~\citep{Sethuraman94}
given by $G(\cdot) = \sum_{j=1}^\infty \psi_j
\delta_{\lambda_j}(\cdot)$, where $\psi_j = v_j \prod_{\ell< j}
(1-v_\ell)$, $ v_j \indep ~ \text{Beta}(1,\alpha)$, $\lambda_j \indep
G_0$ and $\delta_{x}(\cdot)$ denotes the Dirac probability measure
concentrated at $x$. This defines a discrete distribution with
probabilities (weights) $\psi_j$ at atoms $\lambda_j$.

Using a single Dirichlet process is sufficient to cluster rankers or
entities, however, we desire a model that allows for clustering on
both. We therefore appeal to two-way clustering methods and focus on
the Nested Dirichlet Process \citep{rodriguez2008nested}.  The Nested
Dirichlet Process (NDP) is a process whose atoms are Dirichlet
processes.  This model has the desirable feature of having two sets of
skill parameters, say $\lambdavec_1$ and $\lambdavec_2$, that are
either drawn from a distribution over the same atoms with the same
weights, or alternatively, from a distribution over different atoms
with different weights. The NDP can be used when all data points,
$x_{ij}$ ($i=1,\dots,n$, $j=1,\dots,n_i$), are observations in their
own right.  As discussed in Section~\ref{sec:model_rank_process} a
typical observation under the Plackett-Luce model is $\vec{x}_i =
\{x_{i1},\dots,x_{in_i}\}$, whence each element $x_{ij}$ only forms
part of a single observation.  The construction of the NDP dictates
that a Dirichlet process would first be assigned to ranker~$i$ before
the skill parameters are then drawn from the DP with probability
proportional to the likelihood of the ranking.  However, there is no
information contained within a single ranking in order to cluster
entities within $\vec{x}_i$.  We require information from numerous
rankers in order to be able to group entities together; therefore we
need to alter the definition of the NDP in order to make it suitable
for ranked data.  We propose the Adapted Nested Dirichlet Process
(ANDP) and this is the topic of the next section.

\section{Adapted Nested Dirichlet Process (ANDP)}
\label{sec:andp}

The Nested Dirichlet Process needs to be adapted before it can allow
for clustering of both the rankers and entities contained within a
collection of rankings.  Under the standard NDP, rankers are assigned
to a distribution before a sample is then drawn for each of them.  The
adaptation we make dictates that we first draw samples from the
distributions (DPs) based on the information contained within numerous
rankings before then assigning these samples to rankers via another
Dirichlet process.  We let $\Lambda$ denote the collection of
\textit{all} skill parameters in the model and use the notation $Q
\sim \text{ANDP}(\alpha,\vec{\gamma},G_0)$ to refer to the
distribution specified by the stick-breaking representation
$G(\Lambda) =\sum_{s=1}^\infty \psi_s
\delta_{\lambdavec_{s}^*}(\lambdavec^*)$, $G'(\lambdavec_{s}^*)
=\sum_{t=1}^\infty \omega_{st} \delta_{\lambda_{st}}(\lambda),$ where
$\omega_{st} = u_{st} \prod_{\ell<t}(1-u_{s\ell})$, $\psi_s = v_s
\prod_{\ell<s}(1-v_\ell)$, $u_{st} \indep \text{Beta}(1,\gamma_s)$,
$v_s \indep \text{Beta}(1,\alpha)$ and $\lambda_{st} \sim G_0$ for
$s,t \in \mathbb{N}$. A graphical representation of the NDP and the
ANDP, based on Figure~1 in \cite{rodriguez2008nested}, is given in the
Supplementary materials.

It is worth noting that the NDP (and thus ANDP) models are usually specified by two concentration parameters; one controls the top level clustering, in our case rankers, the second corresponds to the lower level clustering, that is, the entity clustering. However, here we introduce an infinite dimensional space for our low level concentration parameters, resulting in the vector $\gammavec = (\gamma_1,\gamma_2,\dots)$. This allows more flexibility within the model as we can handle scenarios where different ranker groups have different levels of entity clustering.

\section{The Bayesian WAND}
\label{sec:wand}  

We are now in a position to define our complete Bayesian model --- the
Weighted Adapted Nested Dirichlet (WAND) process mixture of
Plackett-Luce models --- the main components of which are
\begin{align*}
\vec{X}_i|\vec{\lambda}_i,w_i &\sim \text{PL}_\text{W}(\vec{\lambda}_i,w_i),
\qquad  i=1,\ldots,n,\\
(\vec{\lambda}_1,\ldots,\vec{\lambda}_n)|Q &\sim~Q\\
Q|\alpha,\gammavec,G_0 &\sim \text{ANDP}(\alpha,\gammavec, G_0),
\end{align*}
that is, we use the weighted Plackett-Luce model (\ref{eqn:plprob_weights}) for our data and the ANDP as the prior distribution.

\subsection{Computation}
There are numerous ways to perform inference for Dirichlet process
mixture models.  In our model the low-level Dirichlet processes are
independent, hence allowing standard methods to be used in order to
perform inference under the WAND.  The majority of methods can be
classified as taking either a conditional or marginal approach, as
summarised in, for example, \cite{papa2008retrospective}.  The
conditional approaches typically use truncation in order to
approximate the infinite-dimensional aspect of the stick-breaking
prior, as pioneered by \cite{ishwaran2001gibbs}.  Avoiding
approximations would be beneficial; the slice and retrospective
samplers of \cite{walker2007sampling} and \cite{papa2008retrospective}
provide methods for achieving this.  However, these methods can suffer
from poor mixing and convergence as they attempt to adequately sample
multimodal posterior distributions.  One solution is the addition of
appropriate label switching moves \citep{papa2008retrospective,hastie2015} though, in general, further empirical
work is needed to determine the number and types of move that give an
effective solution.

For these reasons we avoid conditional methods and instead implement a
marginal sampler. These samplers typically involve a P\'{o}lya urn
scheme and marginalise over the infinite-dimensional distribution
\citep{escobar1995bayesian,maceachern1998estimating}, and
thereby avoid the need for approximations.  Algorithm~8 of
\cite{neal2000markov} is one such sampler; this algorithm has been
shown to be one of the most efficient sampling methods for Dirichlet
Process mixtures; see, for example, \cite{papa2008retrospective}.
Also there is no need for additional label switching moves.
Efficiency is achieved by the algorithm only performing updates for
the unique components which are currently assigned to an observation.
Each observation is then assigned to either a component which is
currently in use or to one of $m$ \textit{auxiliary} components which
are drawn from the base distribution.  This algorithm is designed to
sample from a single DP mixture; we therefore propose a nested version
which will enable inference to be performed under the WAND.

\subsubsection{Prior specification}
\label{sec:prior_spec}

Specifying suitable prior distributions is a problem well discussed
within the Bayesian literature~\citep{BernardoS94}.  Here our choice
of priors is mainly for mathematical convenience, however, we believe
they are sufficiently general to portray most prior beliefs.  The
skill parameters $\lambda_k > 0$ must be strictly positive, therefore
a suitable choice for the base distribution is a gamma distribution
with mean $a/b$, that is $G_0 = \Ga(a,b)$. Without loss of generality
we set $b=1$, since $b$ is not likelihood identifiable.  The latent
ranker reliability variables,
$\vec{w}=(w_1,\ldots,w_n)$, are binary indicators and thus we choose $w_i \indep
\text{Bern}(p_i)$ with $p_i \in (0,1]$ for $i=1,\dots,n$.

The DP concentration parameters can of course be fixed constants but
making such choices can be difficult.  Instead we take $\alpha \sim
\Ga(a_\alpha, b_\alpha)$ and $\gamma_s \indep \Ga(a_\gamma, b_\gamma)$
for $s \in \mathbb{N}$ \textit{a priori} which allows posterior
realisations for the concentration parameters to be sampled as in
\cite{escobar1995bayesian}; full details are provided within
Section~\ref{sec:alg_outline}.

\subsubsection{Latent variables}

Before we can detail how posterior computation is achieved we must first introduce some latent cluster indicators.
We introduce $\vec{c} = (c_1,\dots,c_n)$ where $c_i = j$ denotes that ranker~$i$ is associated with parameter vector $\lambdavec_j$. 
Furthermore we require additional indicators to denote the clustering
within each parameter vector; we let $d_{ij} = \ell$ denote that
entity $j$ within parameter vector $i$ is allocated to entity cluster $\ell$.
We shall let $D$ refer to the full collection of the entity cluster indicators.
The value of the skill parameter assigned to entity $j$ from ranking $i$ is therefore
given by $\lambda_{c_i, d_{c_i,j}}$.

The form of the weighted Plackett-Luce likelihood does not admit
conjugate Bayesian inference.  However using latent variables proposed
by \cite{caron2012efficient} gives a semi-conjugate update of the
skill parameters and thereby a Gibbs sampler. This algorithm benefits
from having no reliance on tuning parameters and, in our experience,
is highly efficient.  The latent variables required here are
\begin{equation}
\label{eqn:latent_variables}
z_{ij}|\D,\Lambda,\cvec,D,\vec{w} \indep \Exp \left(\sum_{m=j}^{n_i}  \lambda_{c_i,d_{c_i,x_{im}}}^{w_i} + \sum_{m \in \mathcal{U}_i}  \lambda_{c_i,d_{c_i,m}} ^{w_i}\right)
\end{equation}
for $i=1,\dots,n$, $j=1,\dots,n_i$.
These latent variables
result in the contribution to the complete data likelihood from
ranker~$i$ being
\begin{equation*}
f(\vec{x}_i,\vec{z}_i|\Lambda,\cvec,D,\vec{w}) = \prod_{j=1}^{n_i}  \lambda_{c_i,d_{c_i,x_{ij}}}^{w_i} \exp \left\{ -\left( \sum_{m=j}^{n_i} \lambda_{c_i,d_{c_i,x_{im}}}^{w_i} + \sum_{m \in \mathcal{U}_i}  \lambda_{c_i,d_{c_i,m}}^{w_i} \right)z_{ij} \right\}.
\end{equation*}

\subsubsection{Algorithm outline}
\label{sec:alg_outline}
We are now in a position to detail the algorithm used for sampling
from the posterior distribution $\pi(\Lambda,Z,\vec{c},D,\vec{w},\alpha,\vec{\gamma}|\D)$
under the WAND model.
We first define $N^r = |\{c_i\}_{i=1,\dots,n} |$  to be the current number of ranker clusters and $N^e_s = |\{d_{sj}\}_{j=1,\dots,K} |$ as the number of entity clusters within ranker cluster $s$.
The state of the Markov chain then consists of $\cvec = (c_i)$,  $D = (d_{sl})$, $\Lambda =(\lambda_{st})$, $Z=(z_{ij})$, $\vec{w} = (w_i)$, $\gammavec = (\gamma_s)$ and $\alpha$ for $s=1,\dots,N^r$, $t=1,\dots,N^e_s$, $i=1,\dots,n$, $j=1,\dots,n_i$ and $l=1,\dots,K$. The updates then proceed as follows.

\begin{itemize}
\item For $i=1,\dots,n$:
Let $q^{\text{r}-}$ be the number of distinct $c_j$ for $j \neq i$ and $h^{\text{r}} = q^{\text{r}-}+m^{\text{r}}$. Label these $c_j$ values in $\{1,\dots,q^{\text{r}-}\}$. If $c_i = c_j$ for some $j \neq i$, draw  $\lambdavec_{c} \indep \DP(\gamma_c, G_0)$  for $q^{\text{r}-}<c \leq h^{\text{r}}$. If $c_i \neq c_j \;\forall \;j \neq i$, let $c_i$ have the label $q^{\text{r}-} +1$, and draw  $\lambdavec_{c} \indep \DP(\gamma_c, G_0)$ for $q^{\text{r}-}+1 < c \leq h^{\text{r}}$.

Draw a new value for $c_i$ from $\{1,\dots,h^{\text{r}}\}$ using the
probabilities
\begin{equation*}
\Pr(c_i=c|\D,\Lambda,Z,\cvec_{-i},D,\vec{w}) = 
\begin{cases}
b\:n_{-i,c}^{\text{r}} \:f(\vec{x}_i,\vec{z}_i|\Lambda,\cvec_{-i},c_i=c,D,\vec{w}), &1\leq c \leq q^{\text{r}-},\\
b\:\frac{\alpha}{m^r} \:f(\vec{x}_i,\vec{z}_i|\Lambda,\cvec_{-i},c_i=c,D,\vec{w}), &q^{\text{r}-} < c \leq h^{\text{r}},
\end{cases}
\end{equation*}
where $n_{-i,c}^{\text{r}}$ is the number of $c_j$ for $j\neq i$ that
are equal to $c$, and $b$ is the appropriate normalising constant.
Change the state to contain only those $\lambdavec_c$ that are now
associated with one or more observations. i.e. let $\Lambda =
(\lambdavec_c : c \in \{c_1,\dots,c_n\})$.

\item Relabel $\vec{c}$ such that $c_i \in \{1,\dots,N^r \}$ for $i=1,\dots,n$.

\item For $s=1,\dots,N^r$, $i=1,\dots,K$: Let $q^{\text{e}-}_s$ be the number of distinct $d_{sj}$ for $j \neq i$ and $h^{\text{e}}_s = q^{\text{e}-}_s+m^{\text{e}}$. Label these $d_{sj}$ values in $\{1,\dots,q^{\text{e}-}_s\}$. If $d_{si} = d_{sj}$ for some $j \neq i$, draw  $\lambda_{d} \indep G_0$  for $q^{\text{e}-}_s<d \leq h^{\text{e}}_s$. If $d_{si} \neq d_{sj} \;\forall \;j \neq i$, let $d_{si}$ have the label $q^{\text{e}-}_s +1$, and draw  $\lambda_{d} \indep G_0$ for $q^{\text{e}-}_s+1 < d \leq h^{\text{e}}_s$.

Draw a new value for $d_{si}$ from $\{1,\dots,h^{\text{e}}_s\}$ using the probabilities
\begin{align*}
\Pr(d_{si}=d|\D,\Lambda,Z,\cvec,D_{-si},\vec{w}) &=  \\
&\hspace{-1cm} \begin{cases}
b\:n_{s,-i,d}^{\text{e}} \: \prod\limits_{i \in R} f(\vec{x}_i,\vec{z}_i|\Lambda,\cvec,D_{-si},D_{si}=d,\vec{w}), &1\leq d \leq q^{\text{e}-}_s,\\
b\:\dfrac{\gamma_s}{m^e} \: \prod\limits_{i \in R} f(\vec{x}_i,\vec{z}_i|\Lambda,\cvec,D_{-si},D_{si}=d,\vec{w}), &q^{\text{e}-}_s < d \leq h^{\text{r}},
\end{cases}
\end{align*}
where $n_{s,-i,d}^{\text{e}}$ is the number of $d_{sj}$ for $j\neq i$
that are equal to $d$, $R = \{ i \; : \; c_i = s \}$ and $b$ is the
appropriate normalising constant. Change the state to contain only
those $\lambda$ that are now associated with one or more observations.
i.e. let $\Lambda = (\lambda_{st} : s = 1,\dots,N^r), \; t \in \vec{d}_s=
\{d_{s1},\dots,d_{sK} \})$.

\item For $s=1,\ldots,N^r$ relabel $\vec{d}_s$ such that $d_{sj} \in
  \{1,\dots,N^e_s \}$ for $j=1,\dots,K$.

\item For $s=1,\dots,N^r$, $t=1,\dots,N^e_s$ sample
\begin{equation*}
\lambda_{st} |\D,Z,\cvec,D,\vec{w} \indep \Ga\left(a + \beta_{st},  1+ \sum_{i=1}^n w_i \sum_{j=1}^{n_i} \zeta_{ij}(s,t) z_{ij} \right),
\end{equation*}
where 
\begin{equation*}
\beta_{st} = \sum_{i=1}^n w_i\;\mathbb{I}(c_i = s) \sum_{j=1}^{n_i}  \mathbb{I}(d_{c_i,x_{ij}} = t) \quad \text{and} \quad \zeta_{ij}(s,t) =  \mathbb{I}(c_i = s) \sum_{m=j}^{n_i}  \mathbb{I}(d_{c_i,x_{im}} = t),
\end{equation*}
is the number of times that the element $\lambda_{st}$ represents an
entity in the rankings which are deemed informative
and the number of times an entity associated
with parameter $\lambda_{st}$ is ranked no higher than $j$th in the
$i$th ranking, respectively. The indicator function $\mathbb{I}(A)$
takes value 1 if $A$ is true and 0 otherwise.

\item For $i=1,\dots,n$, $j=1,\dots,n_i$ sample
\begin{equation*}
z_{ij}|\D,\Lambda,\cvec,D,\vec{w} \indep \Exp \left(\sum_{m=j}^{n_i}  \lambda_{c_i,d_{c_i,x_{im}}}^{w_i} + \sum_{m \in \mathcal{U}_i}  \lambda_{c_i,d_{c_i,m}} ^{w_i}\right).
\end{equation*}
\item For $i=1,\dots,n$, sample $w_i$ from the discrete distribution given by
\begin{align*}
\Pr(w_i=1|\D,\Lambda,Z, \cvec,D, \vec{w}_{-i}) &\propto p_i \; f(\vec{x}_i, \vec{z}_i|\Lambda,\cvec,D,\vec{w}_{-i}, w_i = 1),\\
\Pr(w_i=0|\D,\Lambda,Z, \cvec,D,\vec{w}_{-i}) &\propto (1-p_i) \; f(\vec{x}_i,\vec{z}_i|\Lambda,\cvec,D,\vec{w}_{-i}, w_i = 0)\\
  &\propto (1-p_i) \exp \left\{ - \sum_{j=1}^{n_i} z_{ij} (K_i -j +1) \right\}. 
\end{align*}
\end{itemize}

Conditional on the prior distribution discussed within
Section~\ref{sec:prior_spec}, 
the concentration parameters can be sampled from mixtures as follows. 

\begin{itemize}
\item Sample 
\begin{equation*}
\alpha | \cdots \sim \pi \: \Ga(a_\alpha + N^r, b_\alpha - \log\eta) + (1-\pi) \: \Ga(a_\alpha + N^r-1, b_\alpha - \log\eta),
\end{equation*}
where
\begin{equation*}
\frac{\pi}{(1-\pi)} = \frac{a_\alpha + N^r - 1}{n ( b_\alpha - \log\eta)}, \qquad \text{and} \qquad \eta | \cdots \sim \text{Beta}(\alpha + 1, n).
\end{equation*}
\item For $s=1,\dots,N^r$ sample
\begin{equation*}
\gamma_s | \cdots \indep \pi_s \: \Ga(a_\gamma + N^e_s, b_\gamma - \log\eta_s) + (1-\pi_s) \: \Ga(a_\gamma + N^e_s-1, b_\gamma - \log\eta_s),
\end{equation*}
where
\begin{equation*}
\frac{\pi_s}{(1-\pi_s)} = \frac{a_\gamma + N^e_s - 1}{K ( b_\gamma - \log\eta_s)}, \qquad \text{and} \qquad \eta_s | \cdots \indep \text{Beta}(\gamma_s + 1, K).
\end{equation*}
\end{itemize}

\subsubsection{Rescaling}
\label{sec:rescaling}
Consider the sum
$\lambda^\dagger=\sum_{s=1}^{N^r}\sum_{t=1}^{N^e_s}\lambda_{st}$ of
all the $N$ unique parameters within the Markov chain.  As discussed
in Section~\ref{sec:model_rank_process}, the Plackett-Luce likelihood
is invariant to scalar multiplication of the parameters and so
$\lambda^\dagger$ is not likelihood identifiable.  Indeed, if we let
$\Lambda^*=\Lambda/\lambda^\dagger$ then
$\pi(\Lambda^*,\lambda^\dagger|\D) =
\pi(\Lambda^*|\D)\pi(\lambda^\dagger)$. \cite{caron2012efficient}
noted that MCMC schemes for Plackett-Luce models can suffer from poor
mixing without the addition of a rescaling step. The idea is to
rescale the parameters so that the posterior distribution of
$\lambda^\dagger$ is the same as its $\text{Ga}(Na,b)$ prior
distribution. This is achieved at each iteration by taking $\Lambda\to
k\Lambda$, where
$k=\lambda^\dagger/\sum_{s=1}^{N^r}\sum_{t=1}^{N^e_s}\lambda_{st}$ and
$\lambda^\dagger$ is a realisation from a $\text{Ga}(Na,b)$
distribution.

\subsection{Simulation studies}
\label{sec:simstudies}

We have examined the performance of the WAND model in various
scenarios via two detailed simulation studies; the results are given
in Section~\ref{sec:sim_studies} of the Supplementary materials.
These simulation studies demonstrate, amongst other things, that the
WAND model is generally capable of (i) distinguishing between
informative and uninformative rankers, (ii) recovering heterogeneous
structure within the data when it is present, and (iii) producing
reasonable inferences about entity preferences within the
heterogeneous structure. The simulation studies also look at the
effect on inferences of having incomplete data, such as top-$M$
rankings instead of complete rankings. Unsurprisingly, as $M$ gets
closer to the number of entities ($K$) the model is more able to
distinguish between informative and uninformative rankers, and
reproduce any heterogeneous structure within the data.
  
The WAND model has a rich structure with multiple layers of parameters
and latent variables and so there is the potential to have problems of
parameter identifiability. This is often manifested by poor mixing of
MCMC samplers; see, for example, \citet{GelfandS99}. The simulation
studies show that our samplers mix well and also provide other
practical insights into this issue.  For example, consider the
standard Plackett-Luce model ((\ref{eqn:plprob_weights}) with $w_i=1$
for $i=1,\ldots,n$) and the ANDP prior.  It is possible that we do not
need to model uninformative rankings via the weights~$w_i$ as they
could instead be described by a ranker cluster with only one entity
cluster.  However, in practice, we find that fitting this basic model
leads to having several additional (spurious) clusters which house the
uninformative rankings. In contrast, when fitting the weighted
Plackett-Luce model, we find that the weights $w_i$ are not only
identifiable but also stop the introduction of these spurious ranker
clusters.  Also, because these uninformative rankings are generally
severely downweighted in the analysis, they have minimal impact on
inferences for both ranker and entity structure and parameters.
However, a possible drawback is that a WAND analysis may be fairly
conservative in estimating the number of ranker clusters as it has a
preference to declare rankers as uninformative ($w_i=0$) rather than
form a new ranker cluster. Again unsurprisingly we find that inference
about the~$w_i$ can be fairly sensitive to changes in the prior
probability~$p_i$ when the number of rankings is relatively modest.
Therefore, based on these simulation studies, we recommend against
being over confident about ranker abilities \textit{a priori}.
Fortunately, however, inferences about the cluster structure of
rankers and entities is fairly robust to changes in the~$p_i$.

\subsection{Model assessment via posterior predictive checks}
\label{sec:post_pred_checks}

An assessment of the fit of a Bayesian model to the data can be
provided by posterior predictive checks; see, for example,
\cite{GelmanBDA3}.  The basic premise is that the observed data should
look like a plausible realisation from the (posterior) predictive
distribution.  Posterior predictive checks for models for ranked data
have received relatively little attention in the literature, although
\cite{mollica2015bayesian} provide some guidance. Here, rather than
focusing on particular test quantities, we aim to directly assess the
full predictive distribution as follows. In general, ranking $i$
contains $n_i$ entities and therefore the predictive distribution for
ranking $i$ is a probability distribution over the set of all $n_i!$
permutations, though this changes to $K_i!/(K_i-M)!$ permutations for
top-$M$ variants ($n_i=M$).  When the number of all permutations is
relatively small, say $\mathcal{O}(10^5)$, we can calculate the full
predictive probability distribution over all relevant permutations.
This is achieved by evaluating the weighted Plackett-Luce probability
\eqref{eqn:plprob_weights} with $\vec{x}_i$ replaced by each of the
$n_i$ permutations, at each iteration of the MCMC scheme. These
probabilities are then averaged over the MCMC iterations to produce a
Rao-Blackwellised estimate of the predictive distribution for
ranking~$i$. The relative position of each observed
ranking/permutation $\vec{x}_i$ within its predictive distribution can
then be used to assess whether the observed data is plausible, and
hence whether the model provides an adequate description of the data.
Thus we can construct a simple diagnostic plot showing, for each
observed ranking, the proportion of permutations that have a
predictive probability less than or equal to that of the observed
ranking. Here a high value of this predictive probability indicates
that the observed ranking is close to the mode of its predictive
distribution, whereas low values indicate that the ranking is in one
of the tails of the distribution, which might suggest some lack of fit
of the model. When the number of all possible permutations is larger
than say $\mathcal{O}(10^5)$, enumerating the predictive distribution
for each observed ranking becomes computationally infeasible.  Instead
we suggest approximating the distribution by using a Monte Carlo based
approach. Here, for ranker~$i$, we sample $L$ rankings from the
predictive distribution at each iteration.  We then collate these
rankings over all MCMC iterations and carry out the above
Rao-Blackwellisation procedure over a set $\U_i$ containing the
observed ranking $\vec{x}_i$ and these unique sampled permutations.
The idea here is that $|\U_i|$ will be much smaller than $n_i!$ (the
number of possible permutations) and also the sampled permutations
in~$\U_i$ will be ones with relatively high predictive probability.
Thus we can construct a ``truncated'' predictive distribution for each
ranker~$i$ by simply assigning zero probability to those permutations
not in $\U_i$ and rescaling the probabilities for those in $\U_i$ to
sum to one. Clearly constructing the diagnostic plot using the
``truncated'' predictive distribution might lead to the observed
rankings having a lower relative position but the plot should still
give some indication of lack of model fit.  Alternatively, the
diagnostic plot could be constructed using a different approximation
to the predictive distribution, namely one in which all permutations
not in $\U_i$ are given the same (non-zero) probability, that is, the
remaining probability is shared out equally amongst these
permutations. We note that, as $L\to\infty$, we recover the true
predictive distribution for both the ``truncated'' and this
approximate predictive distributions.

\section{Real data analyses}
\label{sec:data}

We now analyse two real datasets which together contain a mixture of
complete and top--$M$ rankings. These datasets have been analysed in
the literature and we compare their conclusions with those obtained
from a WAND analysis. We also give an analysis of a dataset containing
partial rankings in the Supplementary materials.

\subsection{Roskam's dataset}
\label{sec:roskam}

We begin by analysing a dataset originally collected in 1968 by
Roskam, more recently studied by~\cite{de2005nonlinear}, and available
in the R package \textit{homals} \citep{de2009homogeneity}. The data consist of rankings
obtained from $n=39$ psychologists within the Psychology Department at
the University of Nijmengen (Netherlands).  Each ranker gives a
complete ranking of $K=9$ sub-areas (entities), listed
according to how appropriate the sub-areas are to their work. The
sub-areas are: SOC - Social Psychology, EDU - Educational and
Developmental Psychology, CLI - Clinical Psychology, MAT -
Mathematical Psychology and Psychological Statistics, EXP -
Experimental Psychology, CUL - Cultural Psychology and Psychology of
Religion, IND - Industrial Psychology, TST - Test Construction and
Validation, and PHY - Physiological and Animal Psychology.

The heterogeneity within these data has been analysed by
\cite{de2005nonlinear} using a non-linear principal component analysis
to detect groupings within the rankings. Their analysis supported the
idea that there are two groups of rankings: one group which favours
the qualitative fields and the other favouring the quantitative fields
of psychology. A homogeneity analysis was later performed by
\cite{de2009homogeneity} which exposed groupings of entities within
the rankings. More recently \cite{choulakian2016globally} performed a
Taxicab correspondence analysis to look at structure both between the
rankings and the entities within ranker groups.  Their results support
the conclusions of \cite{de2005nonlinear} and suggest that the
psychologists comprise two homogeneous groups with 23 and 16 members
respectively. Within the larger ranker group they obtain the entity
clustering \{MAT, EXP\} $\succ$ \{IND, TST\} $\succ$ \{PHY, SOC, EDU\}
$\succ$ CLI $\succ$ CUL, where $\succ$ means ``is preferred to'', and
quantitative areas of psychology appear to be preferred.  The
corresponding clustering of entities for the other ranker group is
\{EDU, CLI, SOC\} $\succ$ \{CUL, MAT, EXP\} $\succ$ \{TST, IND\}
$\succ$ PHY, and here qualitative areas of psychology appear to be
preferred.  They also conclude that the larger ranker group is
somewhat more homogeneous than the smaller group.

We now use our WAND model to investigate subgroup structure in these
data and take our prior specification for the base distribution and
concentration parameters to be $a=1$ and $a_\alpha=b_\alpha=1$,
$a_\gamma=b_\gamma=3$, as suggested in the Supplementary Materials.
These data contain orderings of individual preferences which we
believe to be informative and so take $p_i=0.75$.  The posterior
distribution is fairly robust to this choice; a sensitivity analysis
is considered in the Supplementary Materials. We report the results
from a typical run of our MCMC scheme initialised from the prior, with
a burn-in of $10$K iterations and then run for a further $1$M
iterations and thinned by $100$ to obtain $10$K (almost)
un-autocorrelated realisations from the posterior distribution.
Convergence was assessed by using multiple starting values, inspection
of traceplots of parameters and the logarithm of the complete data
likelihood, and standard statistics available in the R package
\textit{coda} \citep{codaRpackage}. The MCMC scheme runs fairly
quickly, with C code on a single thread of an Intel Core i7-4790S CPU
(3.20GHz clock speed) taking around $5$~minutes.

Table~\ref{tab:roskam_rank_clust} shows both the prior and posterior
distribution for the number of ranker clusters. The data clearly have
been informative and suggest that it is likely that there are between
two and four ranker groups, with two groups being most plausible.
Note that there is almost no posterior support to suggest there is a
single (homogeneous) ranker group and so an aggregate ranking from
this dataset may be misleading.
\begin{table}[!tbp]
\begin{center}
\setlength\tabcolsep{3.3pt}
\small
\begin{tabular}{l|*{8}{c}}
& 1 & 2 & 3 & 4 & 5 & 6 & 7 & $\geq 8$ \\
\hline
Posterior& 0.00 & 0.43 & 0.33 & 0.16 & 0.06 & 0.02 & 0.00 & 0.00\\
Prior & 0.20 & 0.18 & 0.16 & 0.13 & 0.10 & 0.08 & 0.05 & 0.10\\
\end{tabular}
\caption{Prior and posterior distribution of the number of ranker
  clusters (to 2 d.p.).}
\label{tab:roskam_rank_clust}
\end{center}
\end{table}
The posterior distribution of the allocation of rankers to ranker
groups is, of course, quite complex.  Rather than attempting to
summarise the ranker allocation through a single summary allocation to
each ranker group, such as the maximum \textit{a posteriori} (MAP)
allocation or the improvements to the MAP allocation proposed by
\cite{Dahl06} and \cite{LauG07}, we prefer to use a graphical summary
of the posterior clustering structure.  Following
\cite{MedvedovicS02}, we summarise ranker heterogeneity using a
dendrogram constructed from an agglomerative clustering of the rankers
based on a dissimilarity matrix~$\Delta=(\Delta_{ij})$.  We take
$\Delta_{ij}=\Pr(c_i\neq c_j|\D)$, the posterior probability that two
rankers ($i$ and $j$) are \textit{not} allocated to the same cluster.
We use the complete linkage method, also known as furthest neighbour
clustering, to cluster the rankers as this tends to produce balanced
cluster sizes and does not suffer from ``chaining''.  The complete
linkage method is fairly straightforward: starting with each ranker in
their own singleton cluster, at each stage of the clustering process
the two most similar clusters are merged to form a new cluster; the
dissimilarity between this merged cluster and the remaining clusters
is defined as the maximum of the pairwise dissimilarities between the
members of the merged cluster and the remaining clusters.  Clusters
are merged one at a time until all rankers are in a single cluster.
A dendrogram provides a visual representation of this agglomerative
clustering procedure, with ``branches'' (horizontal lines) drawn
between clusters that are merged; the heights of the horizontal lines
represent the dissimilarity at which the clusters are merged.  Further
details on hierarchical clustering and dendrograms can be found, for
example, in \cite{everitt2011cluster}.

\begin{figure}
\begin{center}
\begin{minipage}[b]{0.5\linewidth}
        \centering
 	    \includegraphics[width=\linewidth, clip, trim=30 17 40 55]{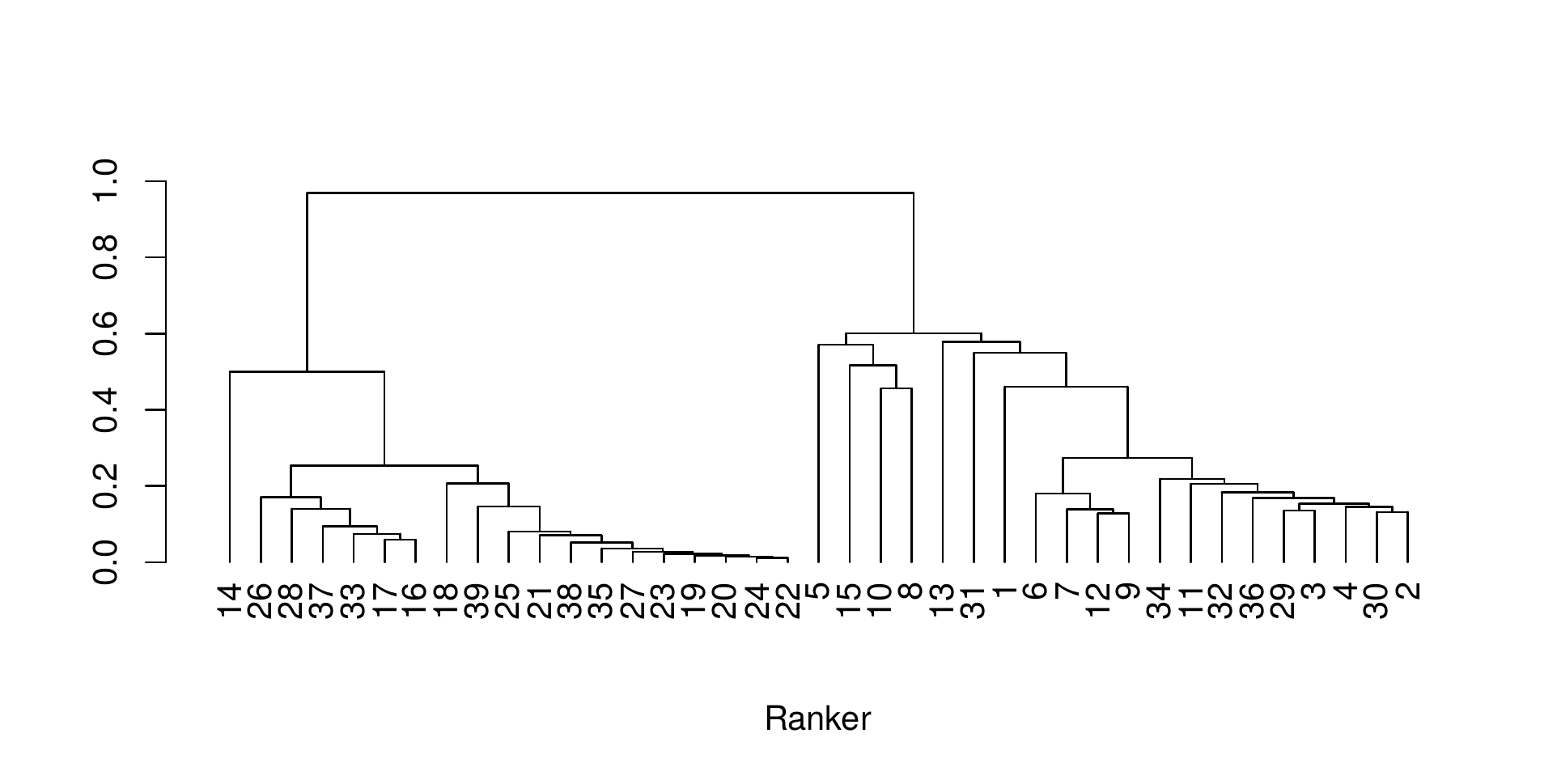}
\end{minipage} 
\hspace*{-0.2cm}
\begin{minipage}[b]{0.5\linewidth}
        \centering
        \includegraphics[width=\linewidth, clip, trim=0 17 30 55]{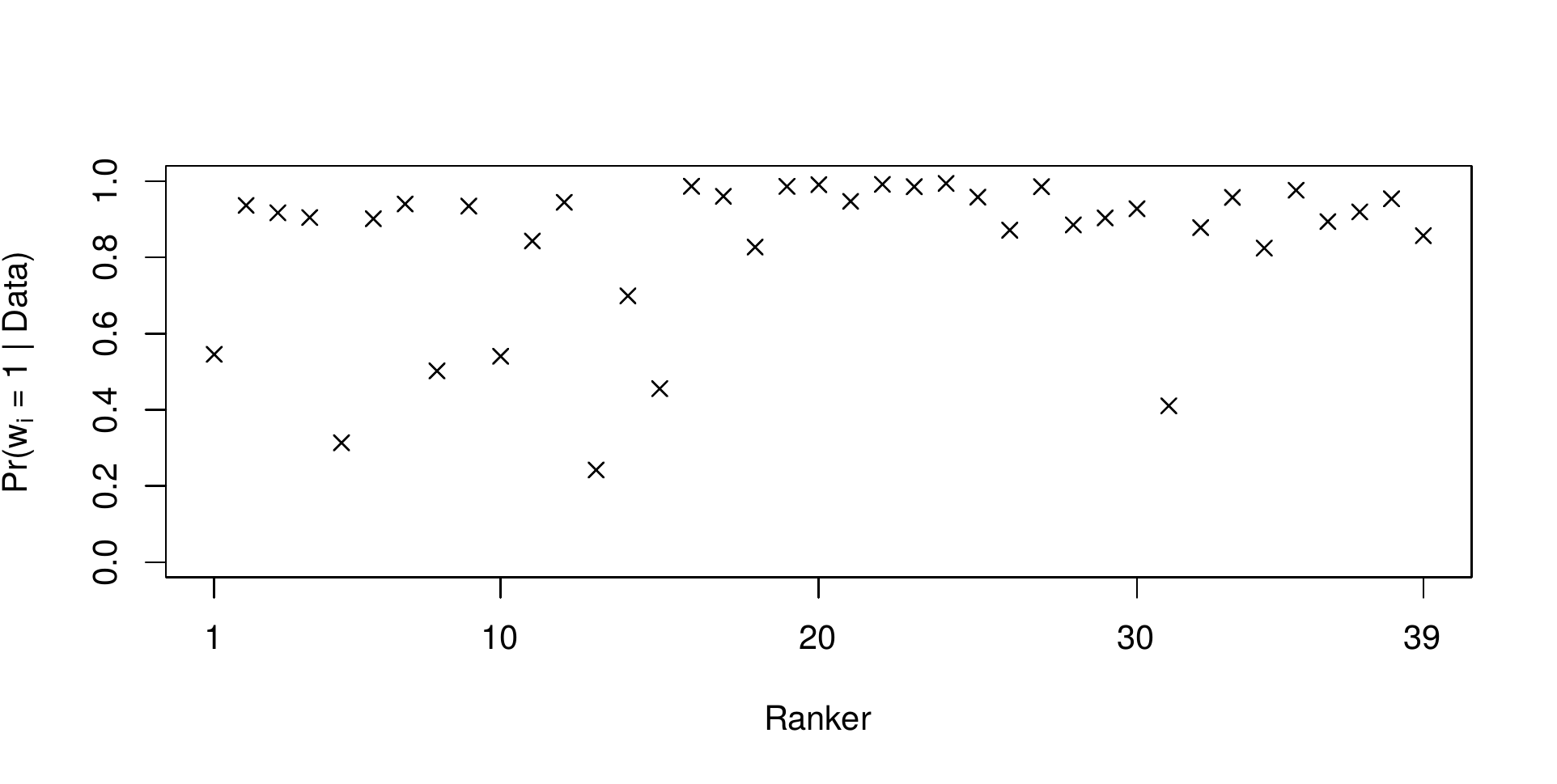}
\end{minipage}
\caption{Roskam's dataset: Dendrogram (left) showing the ranker cluster structure along with the posterior probability, $\Pr(w_i=1|\D)$, for each ranker $i$ (right).}
\label{fig:psy_rank_dend_weight}
\end{center}
\end{figure}

Figure~\ref{fig:psy_rank_dend_weight} shows the dendrogram of rankers
along with the posterior probability that each ranker is informative.
The dendrogram shows that rankers 22 and 24 are the first to be merged
(at height 0.01) and can therefore be considered the most similar pair
of rankers; the posterior probability that these rankers are clustered
together is around $1-0.01=0.99$.  The dendrogram also provides useful
information about the pairwise clustering probabilities between groups
of rankers. For example, the horizontal line connecting rankers 3 and
29, to rankers 2, 4 and 30 indicates that no pair of rankers from the
set \{2, 3, 4, 29, 30\} have a posterior probability of being
clustered together of less than $1-0.15=0.85$. Also note that all
rankers merge at height 0.97, and so the most dissimilar pair of
rankers are clustered together with fairly low posterior probability
of~$0.03$.  Overall this is indicative of fairly strong heterogeneity
in ranker preferences and is consistent with the posterior
distribution in Table~\ref{tab:roskam_rank_clust}.  We note that the
data are consistent with most rankers being informative as
$\Pr(w_i=1|\D)\geq 0.8$, an increase from their prior probabilities
($p_i=0.75$). Also the rankers whose probabilities have decreased
(rankers~$1,5,8,10,13,14,15,31$) are those with (slightly) different
preferences and hence late to join the right-hand cluster in the
dendrogram.

\begin{figure}[t]
\begin{center}
  \includegraphics[width=0.5\linewidth, clip, trim=5 17 30 55]{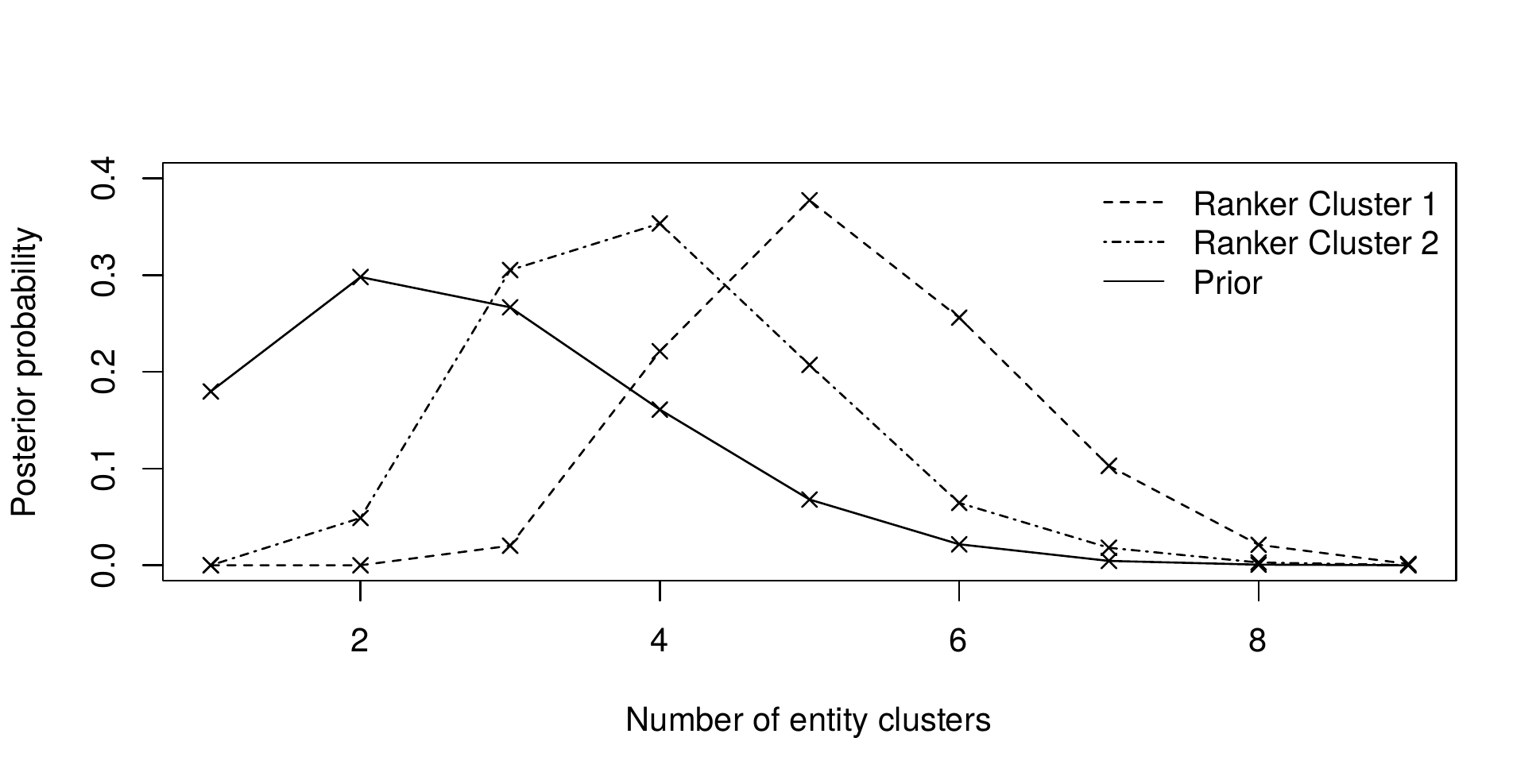}
  \caption{Prior and marginal posterior densities for the number of
    entity clusters within each ranker cluster (conditional on two
    ranker clusters).}
\label{fig:psy_entity_clust_poly}
\end{center}
%\end{figure}
\bigskip
%\begin{figure}
\begin{center}
\begin{minipage}[b]{0.5\linewidth}
        \centering
        \includegraphics[width=0.98\linewidth, clip, trim=30 35 40 50]{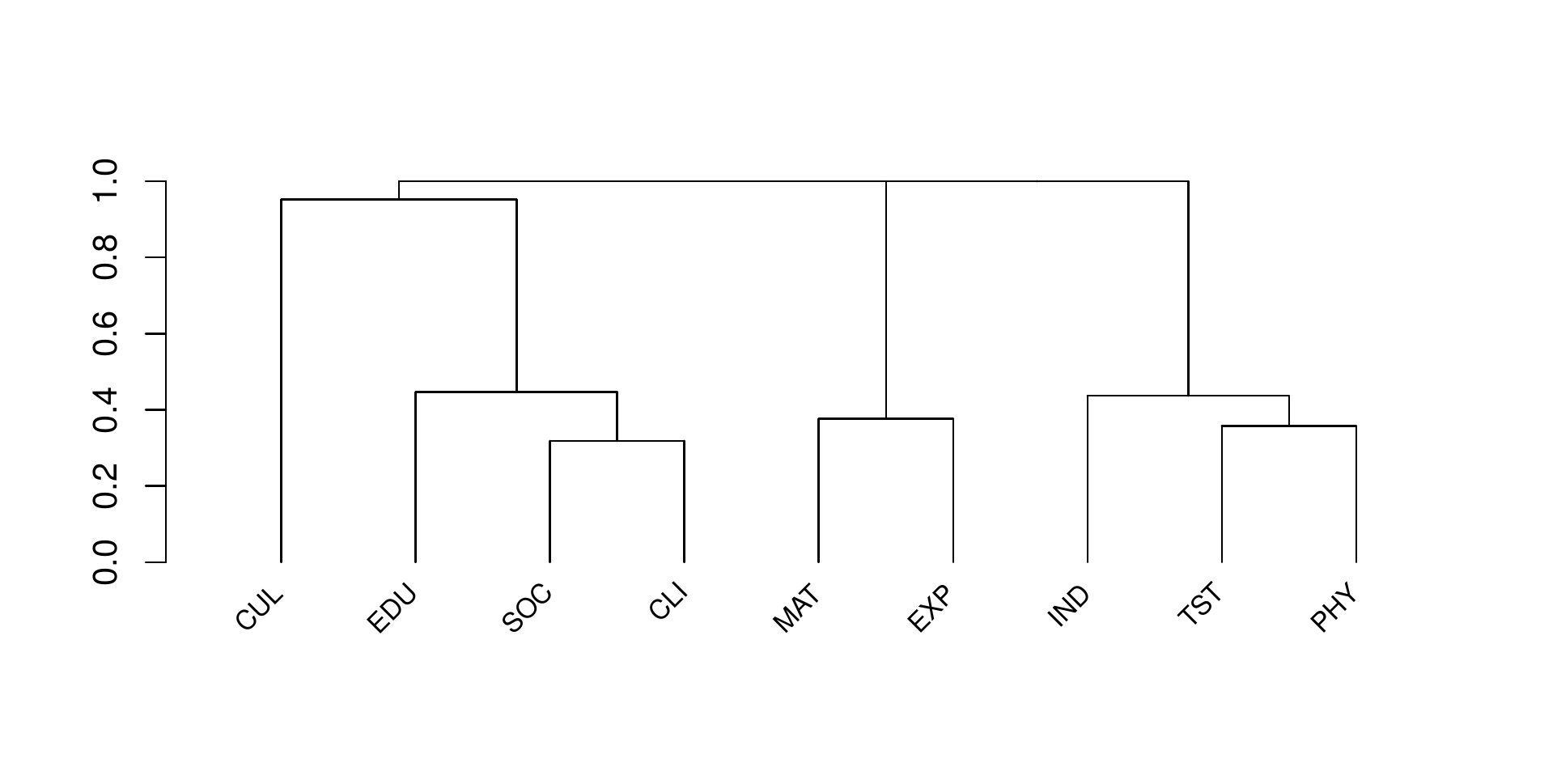}
\end{minipage} 
\hspace{-0.2cm}
\begin{minipage}[b]{0.5\linewidth}
        \centering
		\includegraphics[width=0.98\linewidth, clip, trim=30 35 40 50]{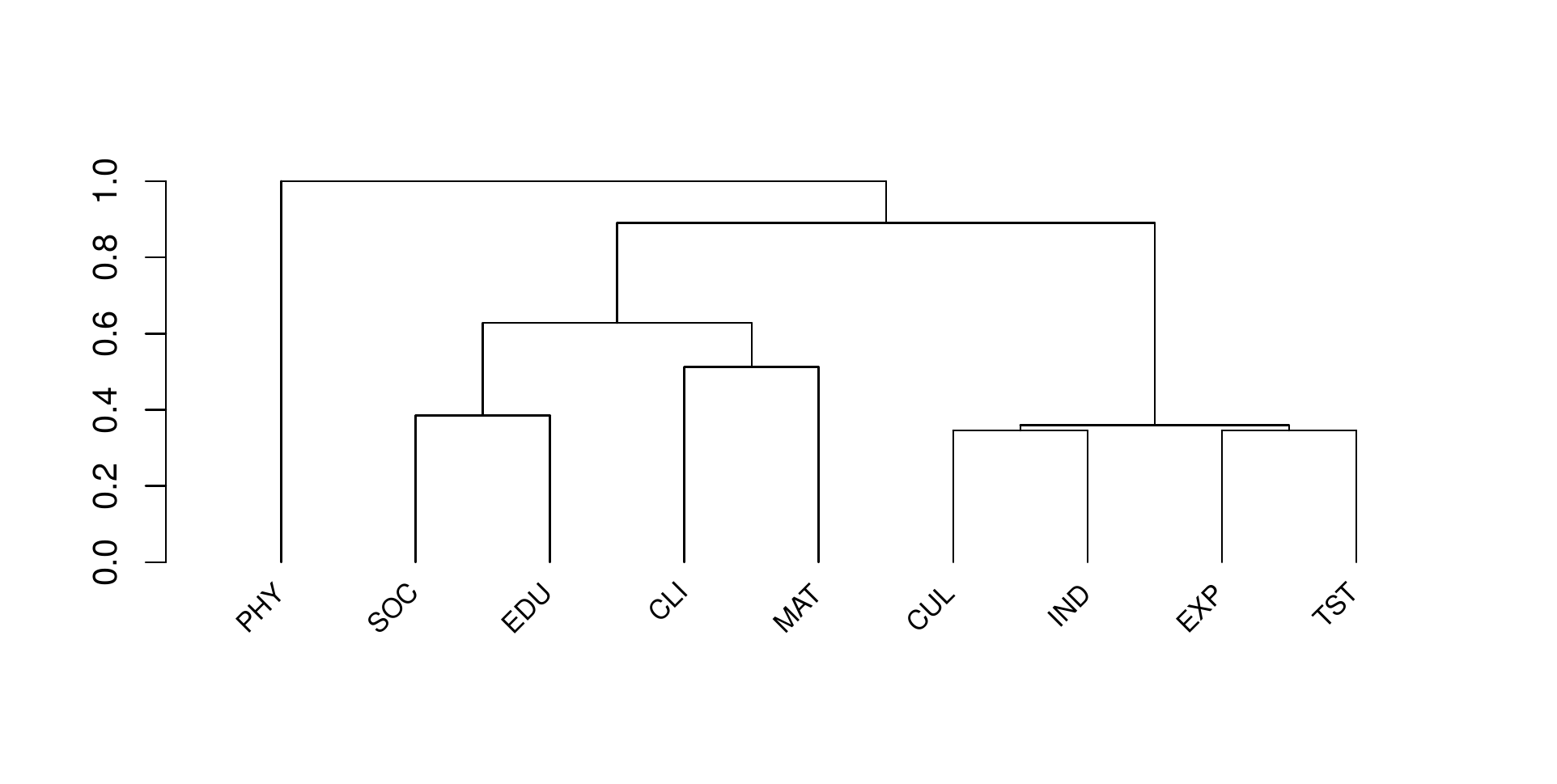}
\end{minipage}
\end{center}
\caption{Roskam's dataset: Dendrograms showing the entity clustering
  structure within ranker cluster 1 and 2 (left and right respectively)
  conditional on two ranker clusters.}
\label{fig:roskam_ent_dend}
\end{figure}

One of the novel features of a WAND analysis is its ability to reveal
subgroup structure of entities within ranker clusters, that is,
highlight entities that are thought to be similar within ranker
groups. We now examine the entity clustering by conditioning on there
being two ranker clusters.  Figure~\ref{fig:psy_entity_clust_poly}
gives the (marginal) posterior distribution for the number of entity
clusters within each ranker cluster, together with the prior
distribution. The dendrograms in Figure~\ref{fig:roskam_ent_dend} show
the entity clustering structure in each ranker cluster, based on
dissimilarity probabilities $\Pr(d_{ij}\neq d_{ij'}|\D)$ of entities
$j$~and~$j'$ within ranker cluster~$i$. We can determine the
membership of the entity clusters by cutting the dendrogram at heights
$\Delta_1\in(0.45,0.95)$ and $\Delta_2\in(0.63,0.89)$ for rankers
groups~1 and~2 respectively and form a preference ordering of these
entity clusters by ordering the marginal posterior means for the skill
parameters $\lambda_{c_id_{c_ij}}$ within each ranker group~$c_i$.
Conditioning on the ranker and entity allocations, we obtain the
ordering \{EXP, MAT\} $\succ$ \{TST, PHY, IND\} $\succ$ \{EDU, SOC,
CLI\} $\succ$ \{CUL\} (with entity cluster means $3.02, 0.72, 0.22,
0.06$) in ranker cluster~1 and \{SOC, EDU, CLI, MAT\} $\succ$ \{CUL,
IND, EXP, TST\} $\succ$ \{PHY\} (with entity cluster means
$1.96,0.82,0.12$) in ranker cluster two.  These entity clusters
(within ranker groups) are similar to those given by
\cite{choulakian2016globally}.  Also if we use the average value of
$\Pr(w_i=1|\D)$ as a measure of homogeneity within a ranker cluster
then we obtain 0.68 and 0.56 for clusters~1 and~2 respectively, which
again agrees with the \cite{choulakian2016globally} conclusion that
ranker cluster~1 is more homogeneous than ranker cluster~2.  Note
that, for this data analysis, we obtain a very similar entity ordering
using marginal posterior means of the skill parameters within each
ranker group (marginal over the distribution of entity clusters); see
Table~\ref{tab:roskam_agg_ranks}. It is interesing to see that the
table suggests that the ranker groups have almost opposite (reverse)
preferences to each other.

\begin{table}[t]
\begin{center}
\footnotesize
\begin{tabular}{c|*{9}{c}}
Ranker & \multicolumn{9}{c}{Rank}\\
cluster & 1 & 2 & 3 & 4 & 5 & 6 & 7 & 8 & 9 \\
\hline
1 & EXP & MAT & TST & PHY & IND & EDU & SOC & CLI & CUL\\
& 3.13 & 2.68 & 0.76 & 0.70 & 0.63 & 0.27 & 0.22 & 0.20 & 0.07 \\
2 & SOC & EDU & CLI & MAT & CUL & IND & EXP & TST & PHY \\
& 1.95 & 1.75 & 1.49 & 1.32 & 0.94 & 0.90 & 0.87 & 0.87 & 0.10 \\
\end{tabular}
\caption{Roskam's dataset: entity rankings by posterior mean within
  ranker cluster (conditional on two ranker clusters). Rank 1
  corresponds to the entity most preferred within each cluster.}
\label{tab:roskam_agg_ranks}
\end{center}
\end{table}

We looked at the sensitivity of the posterior distribution (and
inferences) to modest changes to the prior distribution; full details
are given in the Supplementary materials. Briefly, the posterior
distribution was fairly insensitive to changes in the index ($a$) of
the gamma base distribution and to changes in the parameters
($a_\alpha$, $b_\alpha$, $a_\gamma$, $b_\gamma$) of the gamma prior
distributions for the concentration parameters. The posterior
distribution was most sensitive to changes in the prior probabilities
$(p_i)$ of rankers being informative. Not surprisingly most affected
by such changes were their posterior equivalents $\Pr(w_i=1|\D)$
though the conclusion of two ranker groups and the membership of these
groups was robust. The allocation of entities to groups (within each
ranker cluster) was also fairly robust, with only a minor change in
the allocation when the $p_i$ were increased from ~$0.65$ to~$0.85$.

\subsubsection{Model assessment via posterior predictive checks}

For Roskam's data, each ranker rates $n_i=K=9$ entities and so the
(posterior) predictive distribution for each ranker contains
probabilities for each of the $9!=362880$ possible permutations.
\begin{figure}[t]
\begin{center}
\begin{minipage}[b]{0.49\linewidth}
        \centering
 	    \includegraphics[width=\linewidth, clip, trim=0 17 28 55]{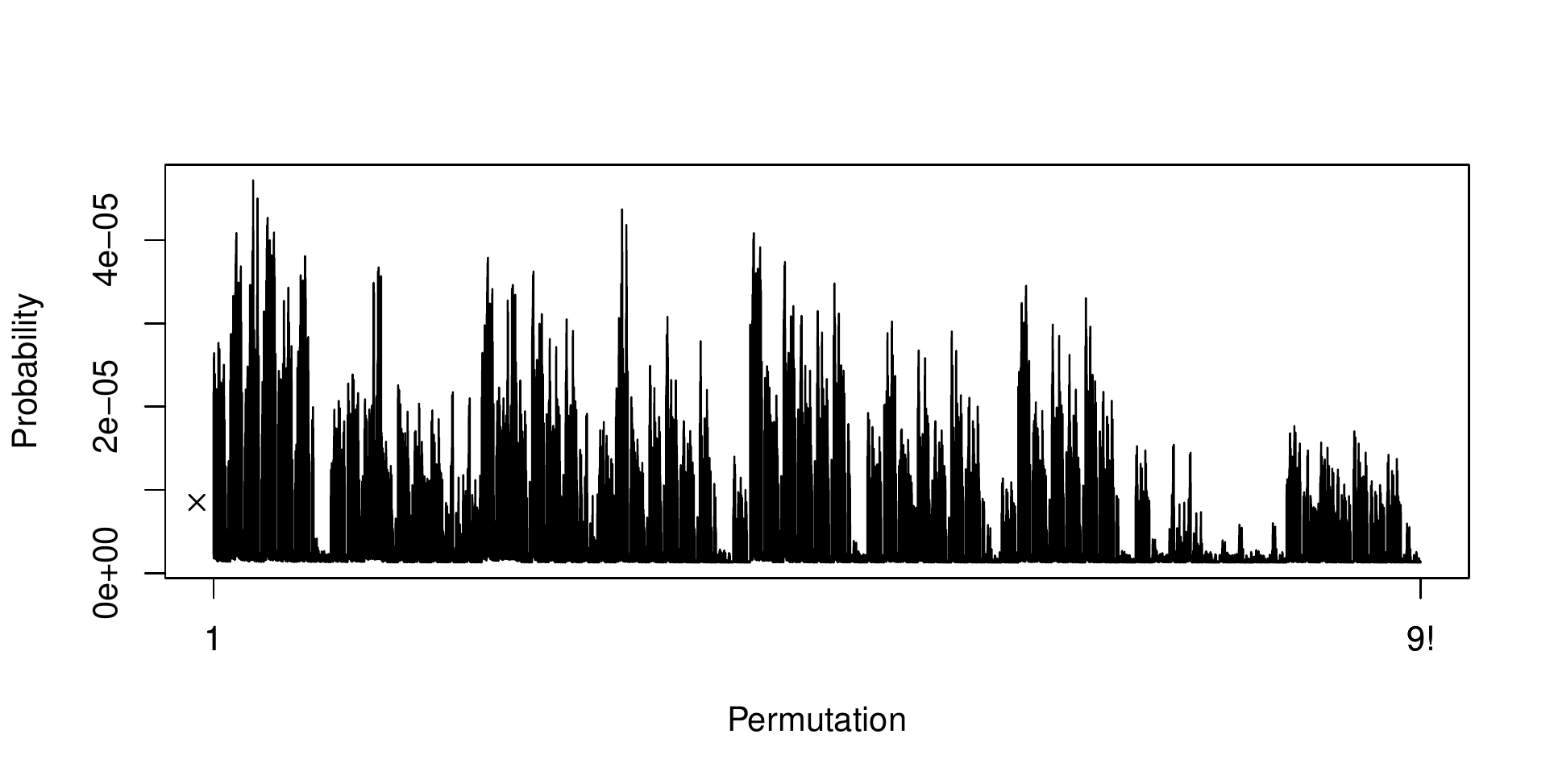}
\end{minipage} 
\hspace*{-0.05cm}
\begin{minipage}[b]{0.49\linewidth}
        \centering
        \includegraphics[width=\linewidth, clip, trim=0 17 28 55]{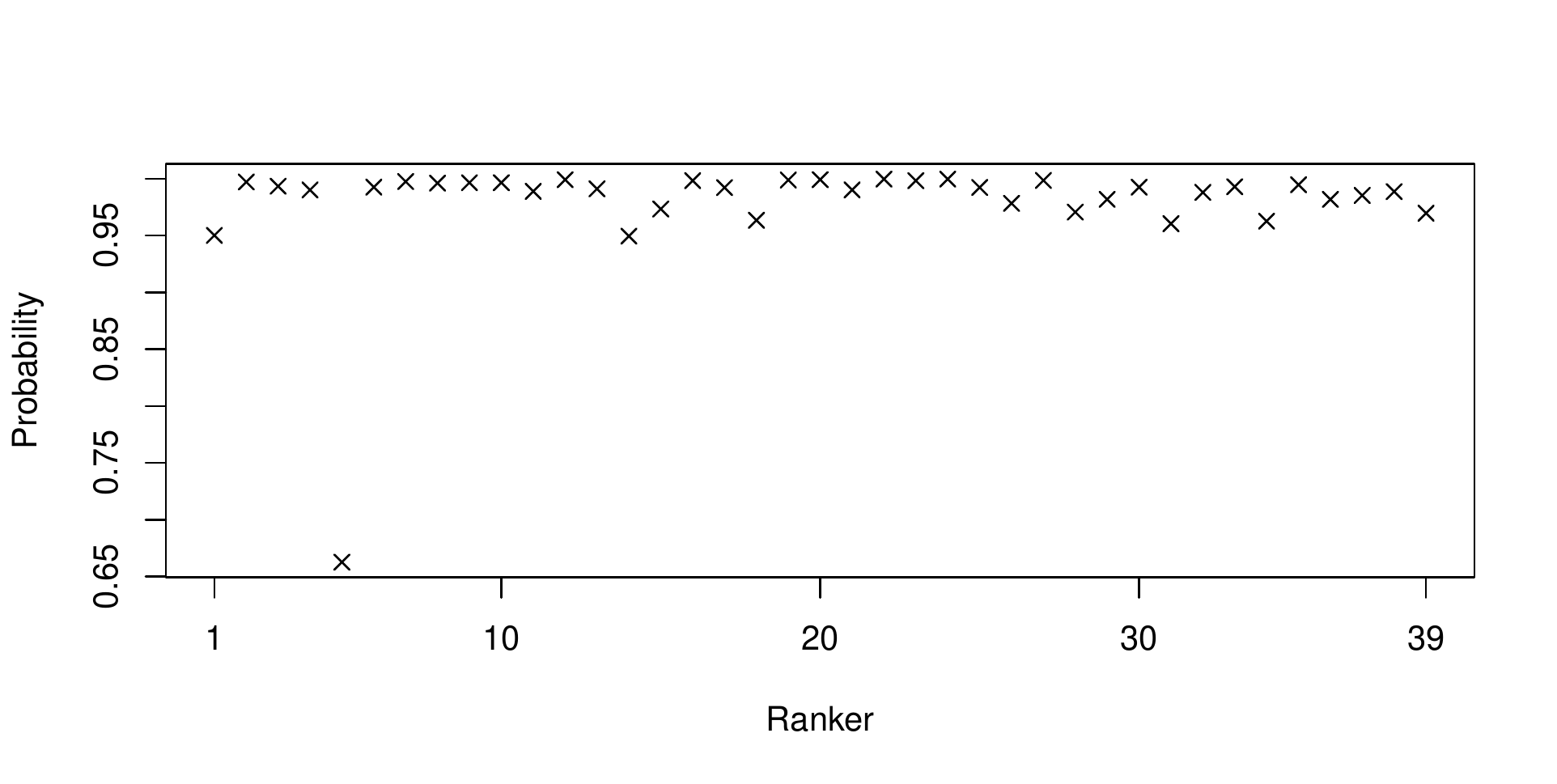}
\end{minipage}
\caption{Roskam's dataset: Posterior predictive distribution (left)
  for ranker~1 (left), the cross shows the posterior predictive
  probability of the observed data~$\xvec_1$. Plot (right) showing the
  diagnostic probability for each ranking (high
    values indicate that an observed ranking is close to the mode of
    its posterior predictive distribution).}
\label{fig:psy_post_pred}
\end{center}
\end{figure}
Figure~\ref{fig:psy_post_pred} (left) shows the predictive
distribution for ranker~1, calculated as described in
Section~\ref{sec:post_pred_checks}. The cross shows the predictive
probability of ranking~$\xvec_1$ provided by ranker~1, and suggests
that this ranking is not unusual, that is, ranking~$\xvec_1$ looks
like a plausible realisation from its predictive distribution. We can
summarise this distribution (and those for other rankers) in a
diagnostic plot (right hand plot in Figure~\ref{fig:psy_post_pred})
that shows the proportion of the $9!$ permutations that have a
predictive probability less than or equal to that of the observed
ranking. The plot shows that all observed rankings are consistent with
their predictive distributions and this suggests that the WAND model
provides a reasonable description of the data.  The most outlying
ranking is that from ranker~5 and, even so, the ranking has a higher
predictive probability than around 65\% of all $9!$ possible rankings.
In Section~\ref{sec:post_pred_checks} we noted that calculating the
full predictive distribution is only computationally feasible
when~$n_i$ is fairly small. Here $n_i=9$ is on the cusp of being
feasible and so we show, in Figure~\ref{fig:psy_post_pred_sim}, the
diagnostic probabilities determined from the truncated and approximate
predictive distributions (for each ranker) using $L=1$, as described
in Section~\ref{sec:post_pred_checks}. The figure confirms that, for
these data, the alternative predictive distributions lead to a similar
conclusion, namely that the observed rankings are consistent with the
WAND model.
\begin{figure}[t]
\begin{center}
\includegraphics[width=0.6\linewidth, clip, trim=0 17 28 50]{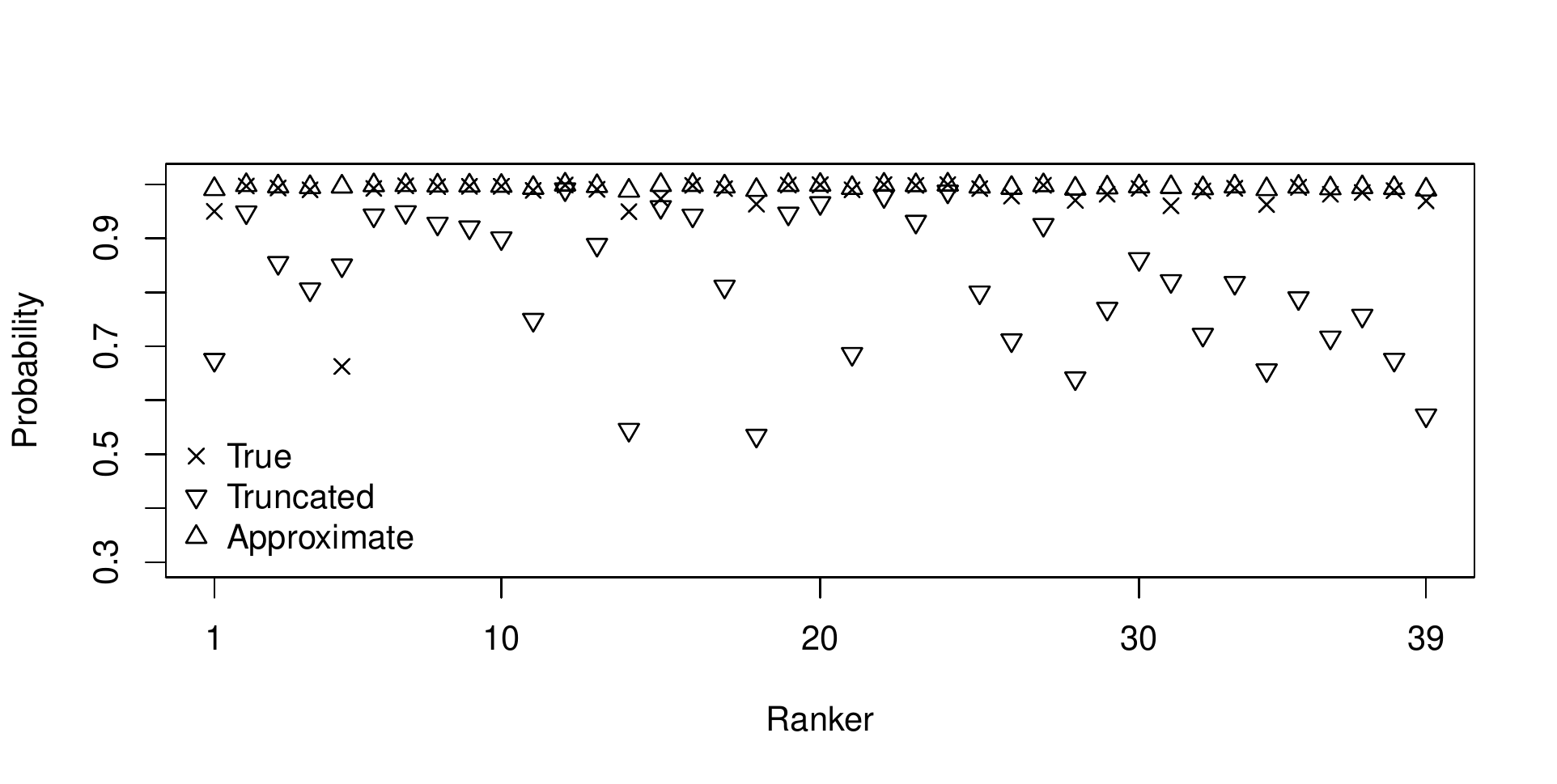}
\caption{Roskam's dataset: diagnostic probabilities for each observed
  ranking $\xvec_i$ calculated using their full, truncated and
  approximate predictive distributions.}
\label{fig:psy_post_pred_sim}
\end{center}
\end{figure}

\subsection{Rankings of NBA teams}
\label{sec:NBA_study}
We now consider another dataset of ranks, studied by
\cite{deng2014rankagg} and involving rankings of NBA (National
Basketball Association) teams. In their paper,
\citeauthor{deng2014rankagg} propose a model (named ``Bayesian
Aggregation of Ranked Data'', BARD) which aims to aggregate rankings
and identify the ``relevant entities''.  Their model also accommodates
the possibility that rankings may not be equally reliable. One
drawback of the BARD model is that it assumes that all rankings come
from a single homogeneous group. We now investigate this assumption.

In 2011/12 the NBA league contained $K=30$ teams (entities) and the
dataset we consider has a ranking of these teams from each of $n=34$
rankers.  The first six complete rankings were obtained from
``professional'' news agencies and the other top--8 complete
rankings obtained from amateurs. Further, each amateur was asked to
classify themselves into one of the following groups: ``Avid fans''
(never missed an NBA game), ``Fans'' (watched NBA games frequently),
``Infrequent watchers'' (occasionally watched NBA games) and ``Not
interested'' (never watched NBA games). Each ranker considered all
teams and so we have $K_i=K$ for $i=1,\dots,n$. The rankers are
numbered as follows: Professionals (1--6), Avid fans (7--12), Fans
(13--18), Infrequent watchers (19--25) and Not interested (26--34).
Therefore we have $n_i=K=30$ for $i=1,\dots,6$ and $n_i=8$ for
$i=7,\dots,n$.  The data are reproduced in the Supplementary
materials.  Further details on how these data were collected can be
found in \cite{deng2014rankagg}.

We now analyse these data using our WAND model and see whether it is
plausible that these rankers are homogeneous or whether the
self-assessed groups behave differently. We take the same prior for
the base distribution ($a=1$) as in the previous example. However, to
reflect weak prior beliefs that there are several ranker groups, we
take $a_\alpha=b_\alpha=3$ in addition to the previous choice for
entities, $a_\gamma=b_\gamma=3$. The prior we choose for each ranker's
ability is based on how much attention they reportedly pay to the NBA,
with professional rankers likely to be most informative, followed by
the Avid fans, then Fans and so on.  We do this by giving the same
$p_i$-value for each ranker in the same ``ability'' group, with
$p_i=0.9$ for professionals, $p_i=0.7$ for Avid fans, $p_i=0.5$ for
Fans, $p_i=0.3$ for Infrequent watchers and $p_i=0.1$ for Not
interested. Of course, in general, information on rankers might well
not be available and so, for comparison purposes we include an
analysis with $p_i=0.5$ in Section~\ref{sec:supp_nba_prior_sens} of
the Supplementary materials. We found that, as seen in the simulation
studies and (other) real data analyses, although the $\Pr(w_i=1|\D)$
were fairly sensitive to changes in the~$p_i$, many other aspects of
the posterior distribution (such as the allocation distributions) were
not.

As in the previous analysis, we report the results from a typical run
of our MCMC scheme initialised from the prior, with a burn-in of $10$K
iterations and then run for a further $1$M iterations and thinned by
$100$ to obtain $10$K (almost) un-autocorrelated realisations from the
posterior distribution. As in the previous analysis, convergence was
assessed by using multiple starting values, inspection of traceplots
of parameters and the logarithm of the complete data likelihood,
together with standard diagnostics available in the R package
\textit{coda}.  Again the MCMC scheme runs reasonably quickly, with C
code on a single thread of an Intel Core i7-4790S CPU (3.20GHz clock
speed) taking just under $18$~minutes.

\begin{figure}
\begin{center}
\begin{minipage}[b]{0.5\linewidth}
        \centering
        \includegraphics[width=\linewidth, clip, trim = 30 17 45 55]{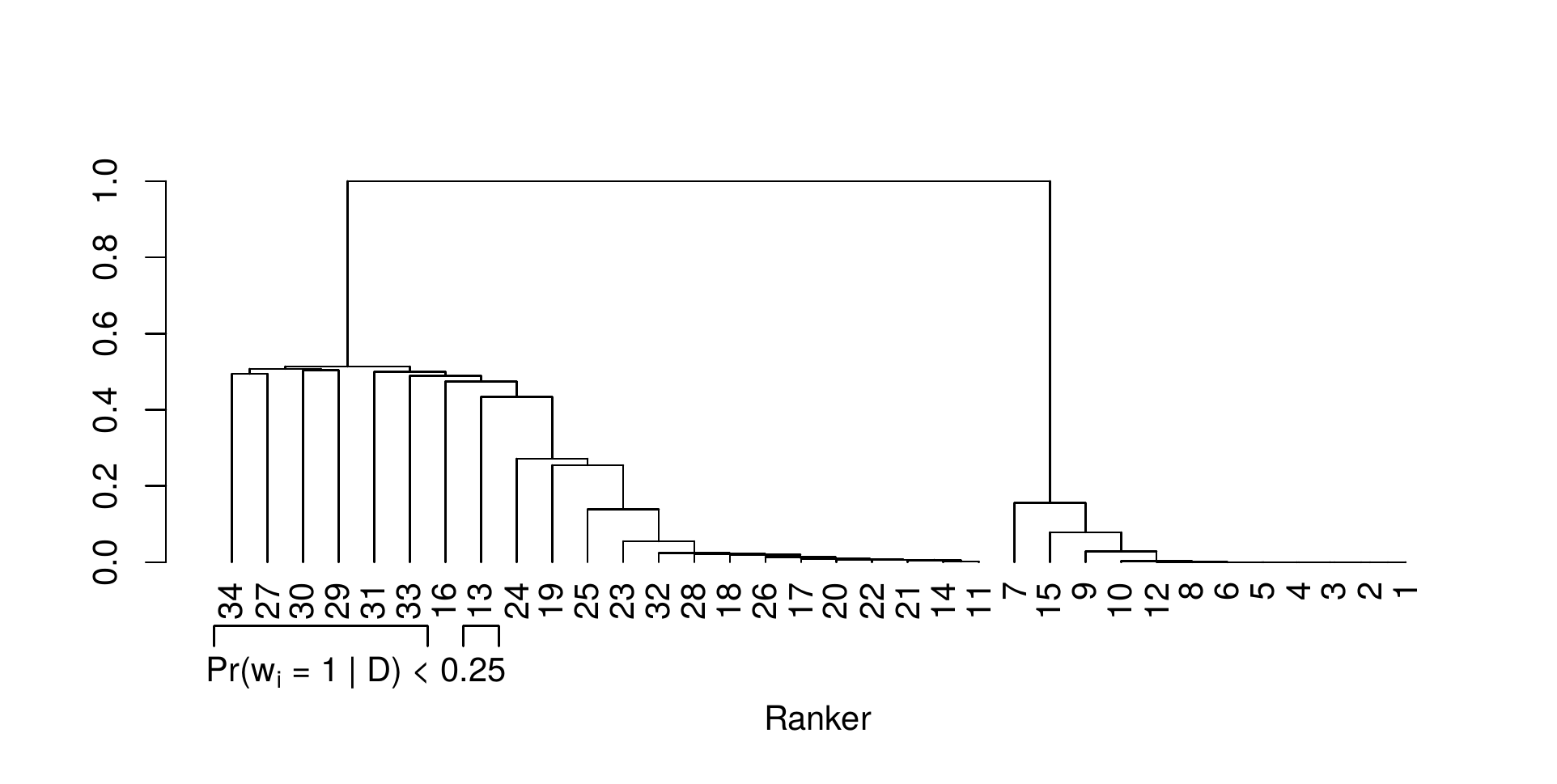}
\end{minipage} 
\hspace{-0.2cm}
\begin{minipage}[b]{0.5\linewidth}
        \centering
        \includegraphics[width=\linewidth, clip, trim = 0 17 30 55]{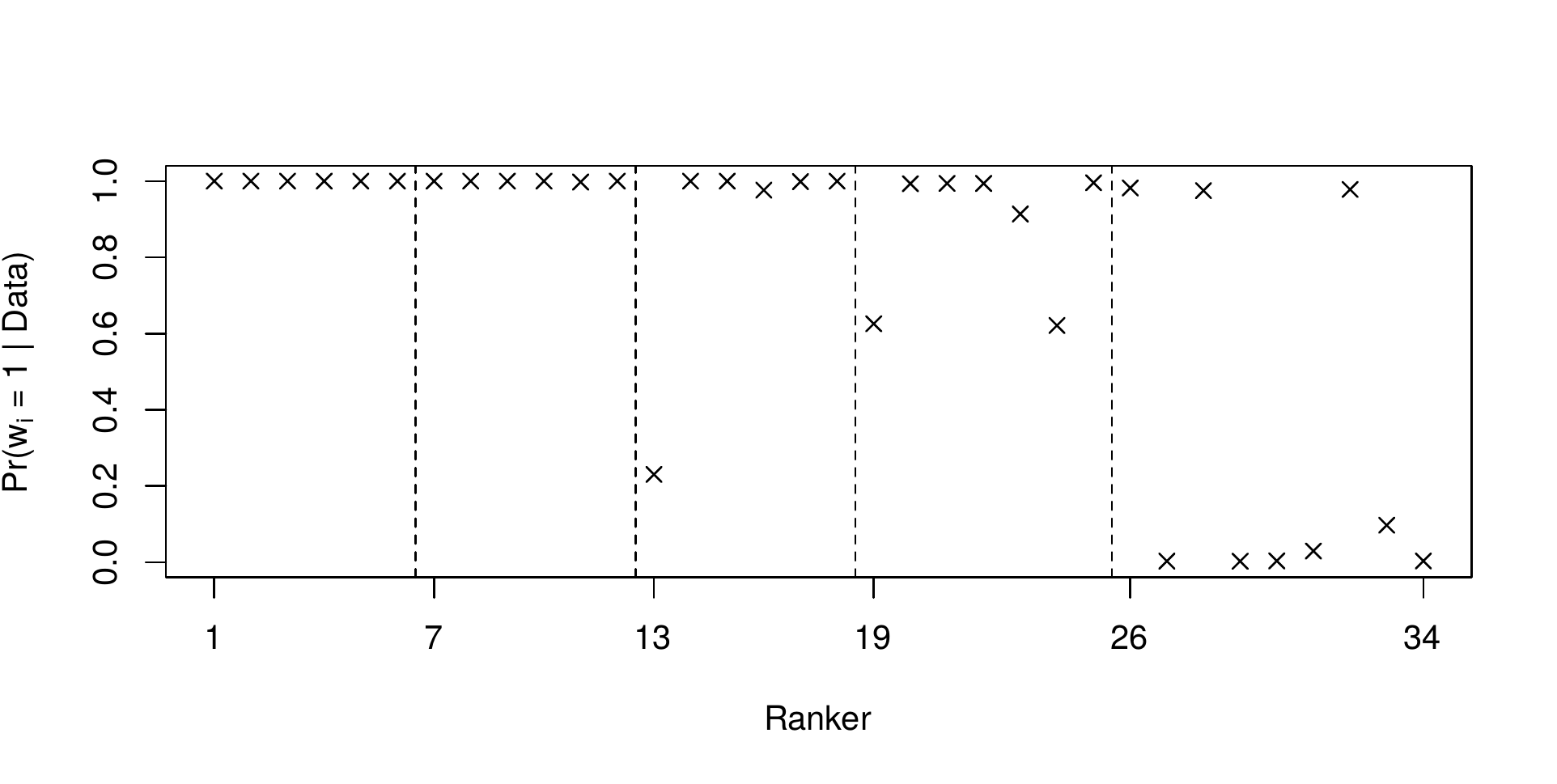}
\end{minipage}
\end{center}
\caption{NBA dataset: Dendrogram (left) showing the clustering
  structure of rankers and highlighting those rankers with
  $\Pr(w_i=1|\D)<0.25$. Plot (right) of the posterior probabilities
  $\Pr(w_i=1|\D)$ for each ranker, with vertical lines separating the
  self-certified groups.}
\label{fig:nba_rank_dend_weights}
\end{figure}

Our analysis of the posterior realisations reveals very little
posterior support for a single homogeneous group of rankers, with most
support for two ranker groups ($\Pr(N^r=1|\D)=0.00$, $\Pr(N^r=2|\D)=0.80$ and
$\Pr(N^r=3|\D)=0.17$).  Figure~\ref{fig:nba_rank_dend_weights} (left)
shows a dendrogram of the posterior clustering structure of rankers
and confirms the conclusion that there are two distinct groups of
rankers: one with rankers 1--10, 12, 15 and the other with rankers 11,
14, 17--26, 28 and 32. Nearly all the other rankers are classed as
uninformative, with $\Pr(w_i=1|\D)<0.25$, except informative ranker~16
who is (roughly) equally likely to be allocated to each cluster; see
Figure~\ref{fig:nba_rank_dend_weights} (right). Note that obtaining a
clustering by using the MAP allocation would be misleading as the MAP
allocation occurs in only 60 of the $10$K iterations in the MCMC chain.
Unsurprisingly, uninformative rankers are typically those who pay less
attention to the NBA, with average values of $\Pr(w_i=1|\D)$ for
rankers in the self-certified groups (from professionals down to the
not interested individuals) of 1, 1, 0.87, 0.88, 0.34 respectively. A
similar conclusion was found under BARD through its ranking quality
parameters; see Figure~8 in \cite{deng2014rankagg}.

\begin{figure}
\begin{center}
  \includegraphics[width=0.5\linewidth, clip, trim=0 17 30 55]{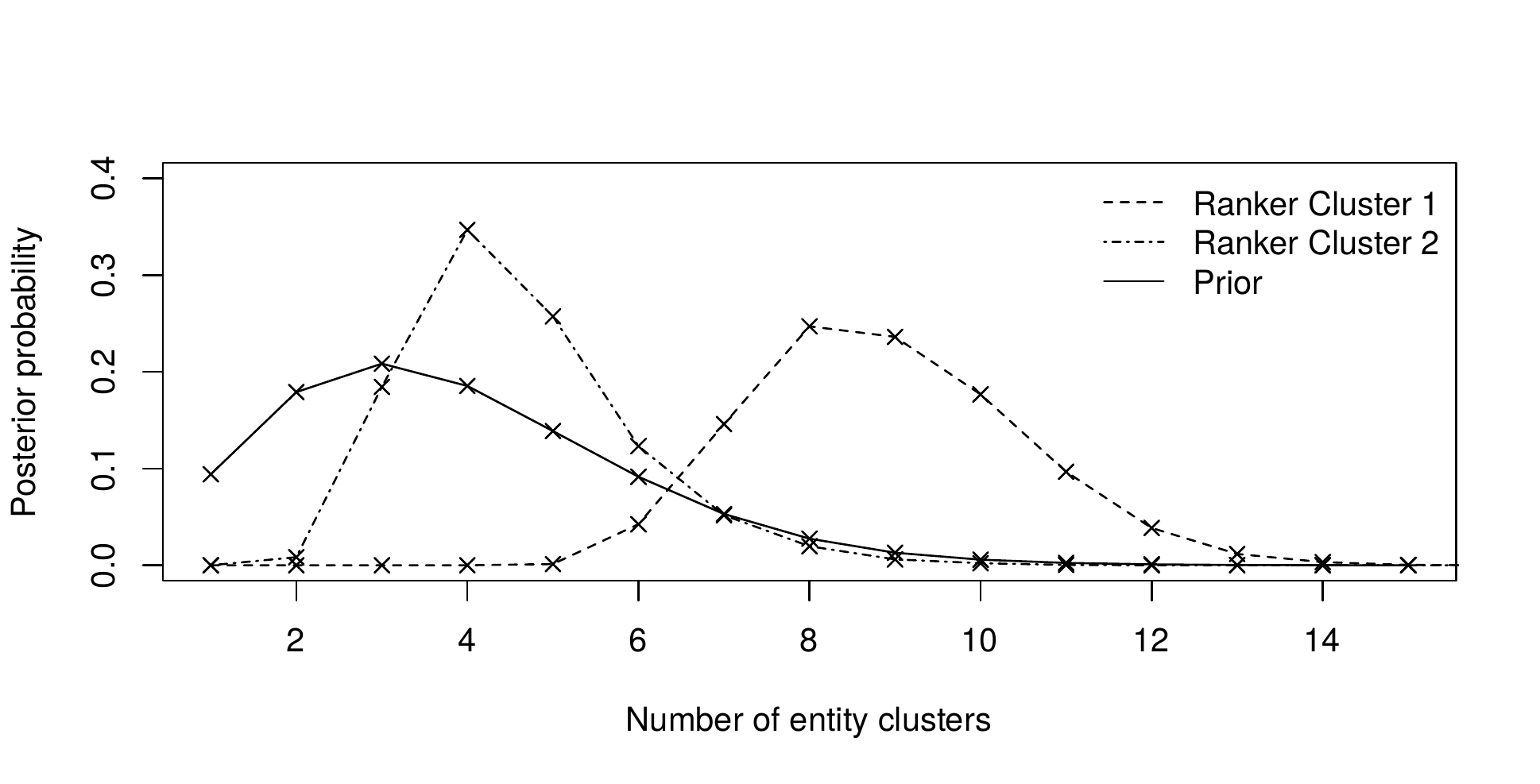}
  \caption{Prior and marginal posterior densities for the number of
    entity clusters within each ranker cluster (conditional on two
    ranker clusters).}
\label{fig:nba_entity_clust}
\end{center}
\end{figure}

\begin{figure}[t]
\begin{center}
\begin{minipage}[b]{0.5\linewidth}
        \centering
        \includegraphics[width=0.98\linewidth, clip, trim=30 35 40 50]{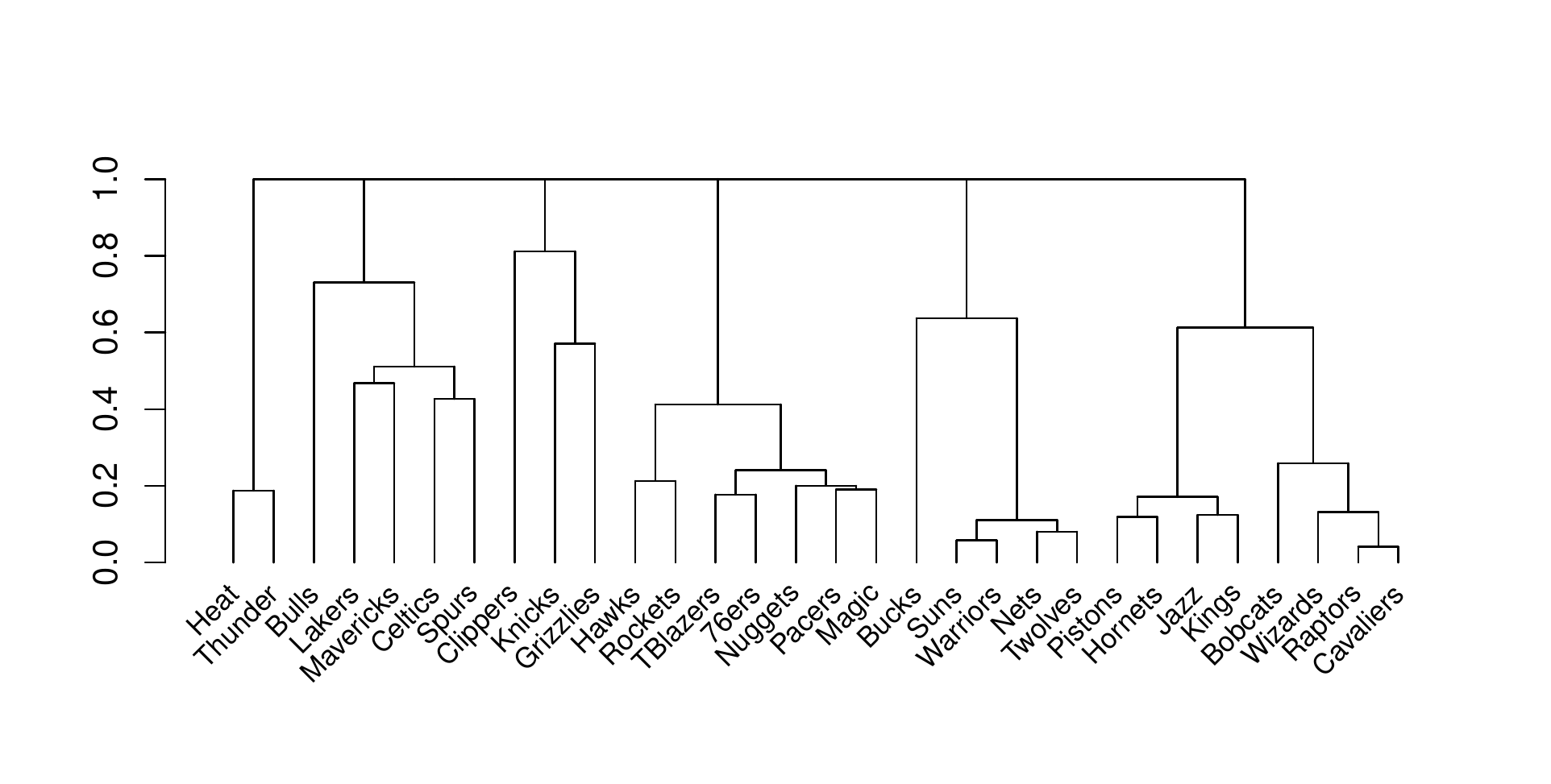}
\end{minipage} 
\hspace{-0.2cm}
\begin{minipage}[b]{0.5\linewidth}
        \centering
        \includegraphics[width=0.98\linewidth, clip, trim=30 35 40 50]{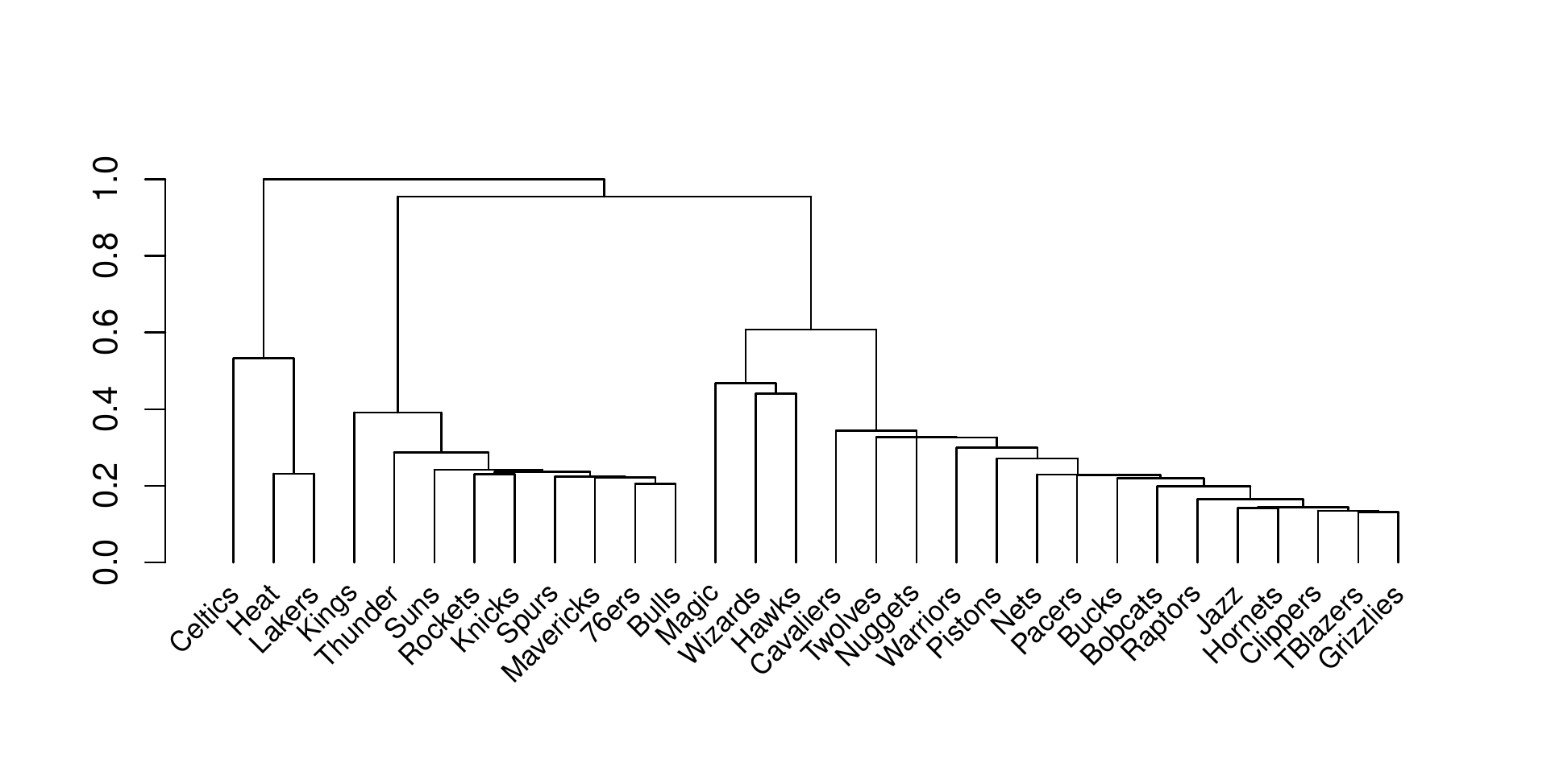}
\end{minipage}
\end{center}
\caption{NBA dataset: Dendrograms showing the entity cluster structure
within ranker clusters~1 and 2 (left and right respectively) conditional on
two ranker clusters.}
\label{fig:nba_ent_dend}
\end{figure}

Figure~\ref{fig:nba_entity_clust} shows the marginal posterior
distribution for the number of entity clusters within each ranker
cluster (conditional on there being two ranker clusters) together with
the prior distribution. The posterior mean number of entity clusters
for ranker clusters~1 and~2 is 8.88 and 4.58 respectively, with
corresponding standard deviations~1.55 and~1.29. These distributions
suggest that rankers within cluster~1 are able to distinguish between
many more entities than those in cluster~2. Again this should come as
no surprise as ranker cluster~2 mainly consists of rankers who
typically pay little attention to the NBA.
The dendrograms in Figure~\ref{fig:nba_ent_dend} show the entity
clustering in each ranker cluster, and suggest that there are six
distinct entity clusters within ranker cluster~1 (taking
$\Delta_1\in(0.81,1)$) and three entity clusters in ranker cluster~2
(taking $\Delta_2\in(0.61,0.95)$). We note that the MAP clustering gives
six and two entity clusters respectively, though there are relatively
few MCMC iterations contributing to the MAP allocation for cluster~2.

It is also of interest to look at the differences in preferences
between the two ranker clusters by examining the within-cluster
aggregate rankings; see Table~\ref{tab:nba_agg} in the Supplementary
materials. As before, these are determined by the marginal posterior
mean for each entity (within each ranker cluster).  The horizontal
lines in this table show the MAP entity clustering described above and
the (quite small) number of occurrences of the MAP is also given.  The
individuals in ranker cluster~1 strongly favour the Heat (entity~1)
and Thunder (2), with the Bulls (10) as the 3rd most preferred team.
Those in ranker cluster~2 also favour the Heat but differ in their
preferences for second and third positions -- here being the Lakers
(6) and Celtics (4). There are many differences in preference
orderings between the ranker clusters, for example, the Thunder and
Bulls appear in positions 11 and 5 in ranker cluster~2. Further
comparison of the BARD and WAND analyses is given within
Section~\ref{sec:nba_bard_comparison_supp} of the Supplementary
materials.
% but we note that the BARD analysis incorrectly assumes that there is only one consensus view from all rankers.

\subsubsection{Model assessment via posterior predictive checks}
\begin{figure}[t]
\begin{center}
\includegraphics[width=0.6\linewidth, clip, trim=0 17 28 50]{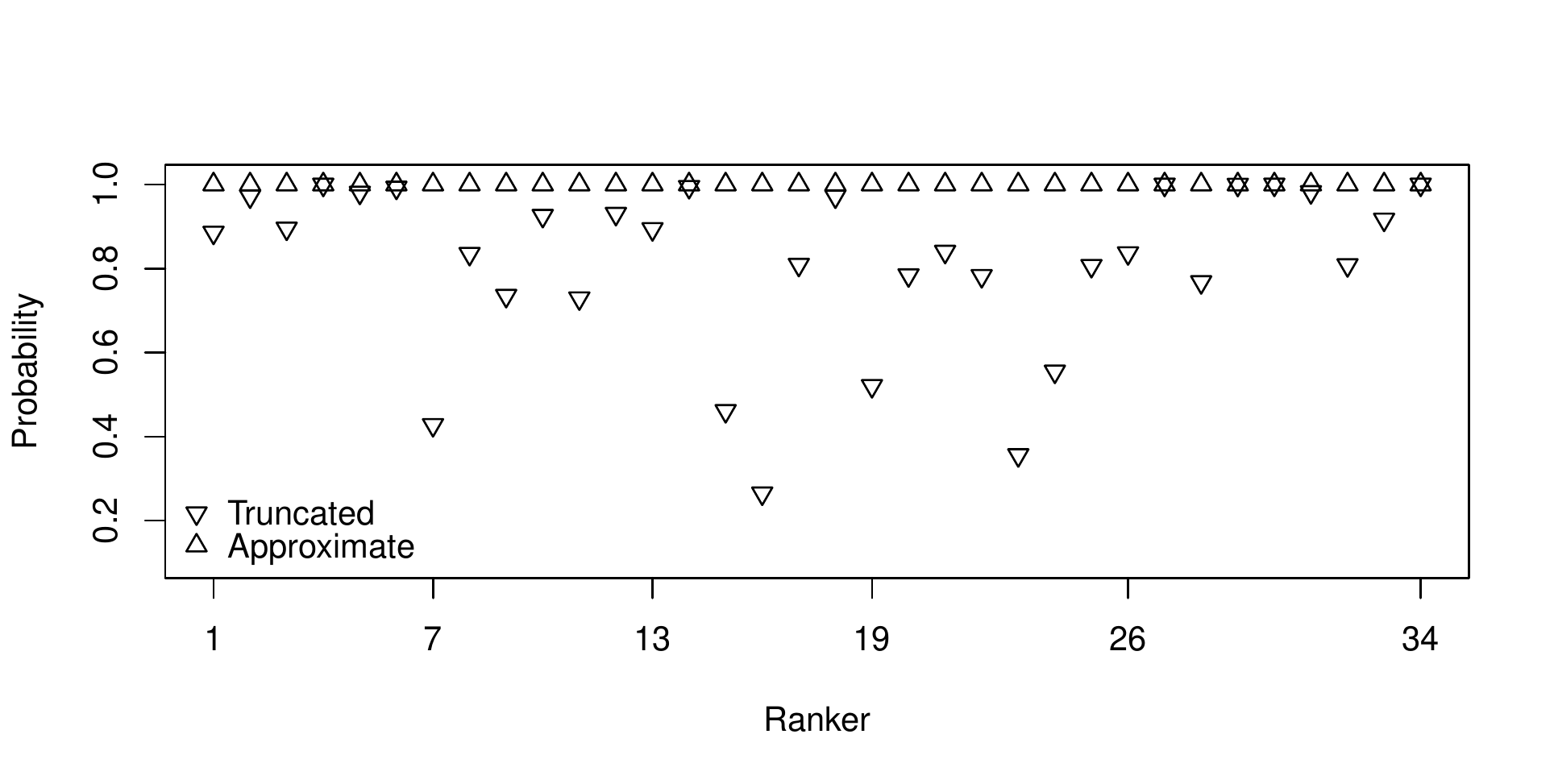}
\caption{NBA dataset: diagnostic probabilities for each observed
  ranking $\xvec_i$ calculated using their truncated and
  approximate predictive distributions.}
\label{fig:nba_post_pred_sim}
\end{center}
\end{figure}

Figure~\ref{fig:nba_post_pred_sim} shows the diagnostic probabilities
for each ranking, calculated from their truncated and approximate
posterior predictive distributions using $L=1$, as described in
Section~\ref{sec:post_pred_checks}.  Note that the probabilities from
the truncated predictive are likely to be fairly conservative as there
are $30!=\mathcal{O}(10^{32})$ possible complete rankings (for rankers
1--6) and $30!/22!=\mathcal{O}(10^{11})$ possible top-8 rankings (for
rankers 7--34), whereas the diagnostic probabilities are determined by
ranking within the (possibly) most likely $\mathcal{O}(10^4)$
permutations.  Overall the figure suggests that the observed rankings
are consistent with the WAND model.

\section{Conclusion}
\label{sec:conc}
In this paper, we have introduced the weighted Plackett-Luce model --
an extension to the standard Plackett-Luce model which includes a
reliability parameter for each ranker which can be inferred from data.
This reliability parameter not only gives us a tool for quantifying
the ability of each ranker but also allows the model to be used in
other purposes such as spam ranker detection. We have found that
inferences under the new weighted Plackett-Luce model are reasonably
robust to the addition of such (spam) rankings whereas the standard
Plackett-Luce model is heavily effected (not reported here).

  The weighted Plackett-Luce model does cater for heterogeneity in
  ranker ability but cannot deal with rankers who express quite
  different preferences.  To handle such scenarios we introduced a
  flexible non-parametric prior distribution, the Adapted Nested
  Dirichlet Process (ANDP), and modelled the ranked data using an
  infinite mixture of weighted Plackett-Luce models with the ANDP as
  the prior distribution (the WAND model). The nested structure of
  this Dirichlet process prior results in our WAND model being capable
  not only of allowing for heterogeneity between rankers but also for
  a flexible grouping structure within the entities themselves.  To
  the best of our knowledge the grouping of entities under the
  Plackett-Luce model has not been considered previously within the
  literature. The (marginal) posterior sampling scheme we outline
  allows for fast and efficient inference under our WAND model.

  The modelling framework described in this paper also allows for
  inferences to be made using only partial information on ranks, such
  as top--$M$ (complete or partial) or partial rankings.  We saw
  through the simulation studies (in the Supplementary materials) that
  reasonable inferences can be made under the WAND model even when
  only limited (partial) information is available.  Although not
  considered here, ties within rankings can easily be accounted for
  within our simulation based inference approach. We have also
  developed posterior predictive checks to help diagnose any problems
  of model adequacy.

  The richness of information in the posterior distribution allows us
  to infer many details of the structure both between ranker groups
  and between entity groups (within ranker groups), in contrast to
  many other (Bayesian) analyses. The high dimension of the posterior
  distribution can make the production of insightful but simple
  summaries quite difficult and we have explored different approaches,
  ranging from conditioning on modal number of groups to adopting a
  classification based on calculations from a dissimilarity matrix
  summary. 

  A fundamental assumption of the weighted Plackett-Luce model is
  that rankers form their rankings under a forward ranking process. In
  certain situations this assumption may not be tenable.  Of course,
  our methods can be adjusted easily to fit a weighted Reverse
  Plackett-Luce model but this model is only appropriate to scenarios
  where the data contain only complete rankings. Currently we are
  studying how to relax the assumption of a ranking process \textit{a
    priori} by considering an infinite mixture of (perhaps weighted)
  Extended Plackett-Luce models \citep{MollicaT14}.

\bibliographystyle{Chicago}
\bibliography{ref_paper}

\end{document}